\documentclass[twocolumn]{aastex62}

\usepackage{longtable} 
\usepackage{rotating, graphicx}
\usepackage{lineno}
\def \oi   	{O\,{\sc i}}

\def \mgii  {Mg\,{\sc ii}}
\def \cii     {C\,{\sc ii}}
\def \si  	{S\,{\sc i}}
\def \ni  	{N\,{\sc i}}
\def \siiv   	{Si\,{\sc iv}}

\def \feii		{Fe\,{\sc ii}}

\def \aia	 	{{\it AIA}}

\def \iris   	{{\it IRIS}}
\def \arcsec 	{\hbox{$^{\prime\prime}$}}
\def \wnth   	{$w_{\rm nth}$}
\def \alfven  {Alfv$\acute{\rm e}$n}

\received{}
\revised{}
\accepted{}

\submitjournal{ApJ}

\shorttitle{IRIS moss variability}
\shortauthors{Testa et al.}

\begin{document}

\title{IRIS observations of short-term variability in moss associated with transient hot coronal loops}
 
\correspondingauthor{Paola Testa}
\email{ptesta@cfa.harvard.edu}

%\author[0000-0002-0405-0668]{Paola Testa}
\author{Paola Testa}
\affil{Harvard-Smithsonian Center for Astrophysics,
60 Garden St, Cambridge, MA 02193, USA}

\author{Vanessa Polito}
\affiliation{Bay Area Environmental Research Institute,
3251 Hanover St, Palo Alto, CA 94304, USA}

%\author[0000-0002-8370-952X]{Bart De Pontieu}
\author{Bart De Pontieu}
\affil{Lockheed Martin Solar \& Astrophysics Laboratory,
3251 Hanover St, Palo Alto, CA 94304, USA}
\affil{Rosseland Centre for Solar Physics, University of Oslo,
P.O. Box 1029 Blindern, NO0315 Oslo, Norway}
\affil{Institute of Theoretical Astrophysics, University of Oslo,
P.O. Box 1029 Blindern, NO0315 Oslo, Norway}

%\author{others}
%\affiliation{}

% The abstract

\begin{abstract}
We observed rapid variability ($\lesssim 60$~s) at the footpoints of transient hot ($\sim 8-10$~MK) coronal loops in active region cores, with the Interface Region Imaging Spectrograph (IRIS). The high spatial ($\sim 0.33$~\arcsec) and temporal ($\lesssim 5$-10~s) resolution is often crucial for the detection of this variability. We show how, in combination with 1D RADYN loop modeling, these IRIS spectral observations of the transition region (TR) and chromosphere provide powerful diagnostics of the properties of coronal heating and energy transport (thermal conduction and/or non-thermal electrons (NTE)). Our simulations of nanoflare heated loops indicate that emission in the \mgii\ triplet can be used as a sensitive diagnostic for non-thermal particles. In our events we observe a large variety of IRIS spectral properties (intensity, Doppler shifts, broadening, chromospheric/TR line ratios, \mgii\ triplet emission) even for different footpoints of the same coronal events. In several events, we find spectroscopic evidence for NTE (e.g., TR blue-shifts and \mgii\ triplet emission) suggesting that particle acceleration can occur even for very small magnetic reconnection events which are generally below the detection threshold of hard X-ray instruments that provide direct detection of emission of non-thermal particles.
\end{abstract}

 
% The main body of the paper

\section{Introduction}
\label{introduction}

The discovery of the high temperature nature of the outer atmosphere of the Sun and solar-like stars more than seven decades ago \citep[e.g.,][]{Grotrian1939,Edlen1943} opened the issue of understanding the origins of the heating of stellar coronae. Though significant progress has been made \citep[e.g.,][]{Klimchuk2006,Parnell2012,Reale2014,Testa2015,Klimchuk2015} the details of the coronal heating are still poorly understood.
Different processes at work in the solar atmosphere are viable candidate as responsible for coronal heating, including dissipation of magnetohydrodynamic \alfven\ waves \citep[e.g.,][]{vanBallegooijen2011,vanBallegooijen2017}, dissipation of magnetic stresses built through random photospheric motions that lead to
braiding of magnetic field lines and causing small scale reconnection events (``nanoflares'', e.g., \citealt{Parker1988,Galsgaard1996,Cargill1996,Priest2002,Gudiksen2005a,Hansteen2015}), as well as processes leading to the formation of spicules and Alfv$\acute{\rm e}$nic waves that can also heat the corona \citep[e.g.,][]{DePontieu2009,DePontieu2010,Bryans2016,DePontieu2017,MartinezSykora2017,MartinezSykora2018}.

Coronal heating likely occurs on small spatial and temporal scales \citep[see e.g., reviews by ][]{Klimchuk2006,Klimchuk2015,Reale2014}, and its signatures are generally difficult to directly detect in the corona because of several factors including the efficient thermal conduction in the coronal, and low emission at high temperature due to the impulsive nature of the heating, with a delayed increase of emission measure with respect to temperature, as well as to non-equilibrium ionization which further decreases the emission at high temperature.

Significant effort has been devoted to revealing the observational signatures of coronal heating, using a variety of imaging and spectral data, in particular focusing on the X-ray and Extreme-ultraviolet (EUV) range. For instance, one of the main predictions of nanoflare heating models is the presence of a hot ($\gtrsim 5$~MK) plasma component. Several studies of imaging observations \citep[e.g.,][]{Reale2009,McTiernan2009,Hannah2016,Grefenstette2016,Ishikawa2019} also in combination with spectroscopic data \citep[e.g.,][]{Ko2009,Testa2011,Testa2012b,Brosius2014,Ishikawa2014,Parenti2017} generally indicate the presence of a hot component with emission measure 2-4 order of magnitude lower than the peak emission (which for AR typically occurs at $\sim 2$-4~MK, e.g., \citealt{Warren2012}). 
Although coronal observations overall suggest significant evidence for hot plasma, the constraints they impose on coronal heating models are not very tight because of the typically large uncertainties due to inherent limitations of the inversion methods to derive the plasma thermal distributions \citep[e.g.,][]{Testa2012c}.

The specific properties of the thermal distribution (emission measure distribution vs. temperature, EM(T)), such as the slopes on both sides of the peak of the EM(T), provide diagnostics of heating frequency and spatial distribution \citep[e.g.,][]{Klimchuk2001,Cargill2004,Testa2005,Bradshaw2012}. Analysis of EM(T) slopes from coronal observations of active regions (ARs) are generally compatible with high frequency coronal heating \citep[e.g.,][]{Warren2012,DelZanna2015}, though low-frequency heating is found to also play a role depending on the evolutionary stage of the AR \citep[e.g.,][]{UgarteUrra2012}.
Additional diagnostics of coronal heating properties are provided for instance by the analysis of time lags imaging observations of AR in different passbands sensitive to different temperature \citep[e.g.,][]{Viall2011,Winebarger2018,Barnes2019} with the Atmospheric Imaging Assembly (AIA, \citealt{Lemen2012}) onboard the Solar Dynamics Observatory (SDO; \citealt{Pesnell2012}). Studies of elemental abundances in ARs at different evolutionary stages \citep{Baker2015,Baker2018}, or during impulsive heating events \citep[e.g.,][]{Warren2016} can also provide clues about heating properties, since the coronal abundances appear to vary with respect to photospheric abundances and this fractionation process is likely related to the heating process \citep[e.g.,][]{Feldman1992,Testa2010,Testa2015}.

Some of these difficulties in revealing direct signatures of nanoflares in the corona can be overcome by searching for those signatures in the lower and cooler layers of coronal structures, i.e., in the transition region (TR) between the cooler photosphere/chromosphere and the hotter corona. The TR of AR loops, also called "moss", is characterized by large gradients of plasma density and temperature over a very narrow layer of few thousand km \citep[e.g.,][]{Fletcher1999,Berger1999,Warren2008}, and is very sensitive to heating. The moss has often been found to be characterized by low temporal variability, which has been interpreted as evidence of high frequency ($\sim$ steady) heating \citep[e.g.,][]{Antiochos2003,Brooks2009,Tripathi2010}. However, this could be at least partly due to the insufficient spatial and temporal resolution of early moss observations. 

Recent observations of moss at higher spatial ($\sim 0.3$-0.4~\arcsec) and temporal ($\lesssim 5$~s) resolution with the High-resolution Coronal Imager (Hi-C, \citealt{Kobayashi2014}) and the Interface Region Imaging Spectrograph (IRIS, \citealt{DePontieu2014}) have indeed shown high variability in the moss, associated with heating of overlying coronal loops to high temperatures ($\sim 5$-10~MK; \citealt{Testa2013,Testa2014}). 
Hi-C observations revealed high variability in the moss on $\lesssim 30$~s scales constraining the energy and duration of impulsive heating events \citep{Testa2013}. The analysis of IRIS spectral observations revealed additional diagnostics of the mechanism of energy transport--thermal conduction (TC) vs.\ non-thermal electrons (NTE)--via modeling of the Doppler shifts observed in the IRIS \siiv\ 1402.77\AA\ line \citep{Testa2014}. These new diagnostics of non-thermal electrons are particularly interesting because: (1) they allow us to reveal indirect signatures of NTE in nanoflares which are generally below the threshold of detectability of hard X-ray telescopes, such as RHESSI \citep{Lin2002}, which directly detect the emission of NTE, and (2) they can constrain the properties of NTEs, such as the low-energy cutoff ($E_C$) of their power-law distributions, which are poorly constrained by hard X-ray spectra because of the overlap of thermal and non-thermal spectra.

These observations of highly variable moss brought forth several questions: how common are these events? what is the frequency of nanoflare events? what is the typical energy released in these nanoflare events? how important are non-thermal particles in small heating events (i.e., for non-flaring active regions heating)? when NTE are present, what are their properties (in particular $E_C$)? 
In the study presented in this paper we manually selected a sample of coronal/TR/chromospheric datasets observing heating of coronal loops associated with short-lived footpoint brightenings to address some of these important issues. We searched the IRIS database for this type of events, though we found a relatively limited number of them because of the difficulty in finding the footpoint brightenings under the slit and the unwieldiness of the manual selection (see \S~\ref{observations}).

A key ingredient for these diagnostics based on IRIS TR spectra are the numerical simulations of nanoflare heated loops. In \cite{Testa2014} we presented exploratory 1D hydrodynamic loops simulations with the RADYN code (\citealt{Carlsson1997,Allred2005,Allred2015}; see \S\ref{sims} for details) to showcase the diagnostic potential of the \siiv\ Doppler shift in impulsive heating events. In \cite{Polito2018} we greatly expanded these initial results and carried out a more thorough exploration of the parameter space and relevant diagnostics. There however we mostly focused on the TR \siiv\ emission, and partly on the \mgii\ chromospheric emission.
Here, in order to further constrain the interpretation of the observed spectral properties, we expanded the analysis of the RADYN simulations both by looking more in detail at the chromospheric properties (\cii\ and \mgii\ triplet) and also expanding the exploration of the parameter space (see \S~\ref{sims}), also combining heating by thermal conduction and NTE.

In this paper, we describe the data selection, and present a detailed analysis of the  \iris\ spectral properties of chromospheric and transition region lines, and their statistical properties across the studied sample (\S~\ref{observations}). We also present some of the properties of the heated coronal loops, but many of the coronal properties are analyzed and discussed in more detail in \cite{Reale2019a}. In Section~\ref{sims} we describe the results of the new RADYN simulations and resulting diagnostics. In Section~\ref{discussion} we discuss the comparison of observed and predicted properties of the chromospheric and TR brightenings and use the simulations as an aid for interpreting the observations. In Section~\ref{conclusions} we draw our conclusions and discuss future prospects.

\begin{deluxetable*}{cccccccccc}
\tablecaption{List of observations showing moss variability.}
\tablehead{
 &  &  & &  \colhead{\iris} &  &  & & \colhead{\aia} & \\ \cline{4-6} \cline{8-10}
\colhead{No.} & \colhead{date and time} & \colhead{AR} & \colhead{OBSID} & \colhead{$t_{\rm exp}$} & \colhead{SG cadence} &  & \colhead{L\tablenotemark{a}} & \colhead{peak $I_{94}$} & \colhead{$\Delta t_{94}$\tablenotemark{b}} \\
 &  &  & &  \colhead{[s]} &  \colhead{[s]} & & \colhead{[Mm]} & \colhead{[DN/s/pix]} & \colhead{[s]}
}
\startdata
1  & 2014-02-23 23:23UT & 11982    & 3860257403 & 4 & 5      & & 15    &  38 & 192 \\
2  & 2014-04-10 02:46UT & 12026    & 3820257403 & 4 & 5      & & 7    &  470 & 648 \\
3  & 2014-09-17 14:45UT & 12166/7 & 3820507453 & 4 & 5.3   & & 30-45  & 120 & 1692 \\
4  & 2014-09-17 17:20UT & 12166/7 & 3820257452 & 4 & 5.3   & & 50-70     &  32 & 2028 \\
5  & 2014-09-18 08:08UT & 12166/7 & 3820257453 & 4 & 5.4   & & 45     &  33 &  480 \\
6  & 2015-01-29 18:29UT & 12268    & 3820257453 & 4 & 5.4   & & 60        &  215 & 984 \\
7  & 2015-11-11 02:47UT & 12450     & 3600104304 & 4 & 3.1   & & 18     & 16 & 672 \\
8  & 2015-11-12 01:38UT & 12450     & 3600104017 & 2 & 12.8 & & 35     & 22 & 612 \\
9  & 2015-12-24 15:21UT & 12473     & 3680086903 & 4 & 5.7   & & 40        & 208 & 1008 \\
10& 2016-01-29 06:23UT & 12488     & 3664251603 & 1 & 2.6  & & 15      & ($<$)10 & 348 \\
\enddata
\label{table_obs}
\tablenotetext{a}{Estimate of projection of the coronal loop length in the plane-of-the-sky}
\tablenotetext{b}{Duration of the coronal event, estimated calculating where $I \ge I_{\rm max}/e$ in the \aia\ 94\AA\ emission averaged over a region including the transient loops.}
\end{deluxetable*}

\section{Observations and Data Analysis}
\label{observations}

In this paper we study a sample of \iris\ and \aia\  datasets showing rapid variability at the foopoints of coronal loops which is caused by the plasma response to impulsive coronal heating. As demonstrated by our previous analysis of one of these events \citep{Testa2014}, temporally resolved spectral data provide valuable diagnostics of coronal heating. Therefore we searched for datasets showing rapid moss variability, at least partly under the \iris\ slit, and followed by increased emission in the hot \aia\ 94\AA\ passband, to confirm that the chromospheric/TR brightening is associated with coronal heating. 

By manually looking through the \iris\ archive we found the events listed in Table~\ref{table_obs}. In Table~\ref{table_obs} we describe the properties of the \iris\ observations: (1) observing program (OBSID) --the selected datasets are either sit-and-stare (i.e., with the slit always at the same location on the Sun, i.e., including solar rotation compensation) or small rasters (2 or 4 step rasters), to ensure high cadence in the spectral data--, (2) integration time ($t_{\rm exp}$), and (3) spectrograph (SG) cadence.

\begin{figure*}
	\centering
    \includegraphics[width=14cm]{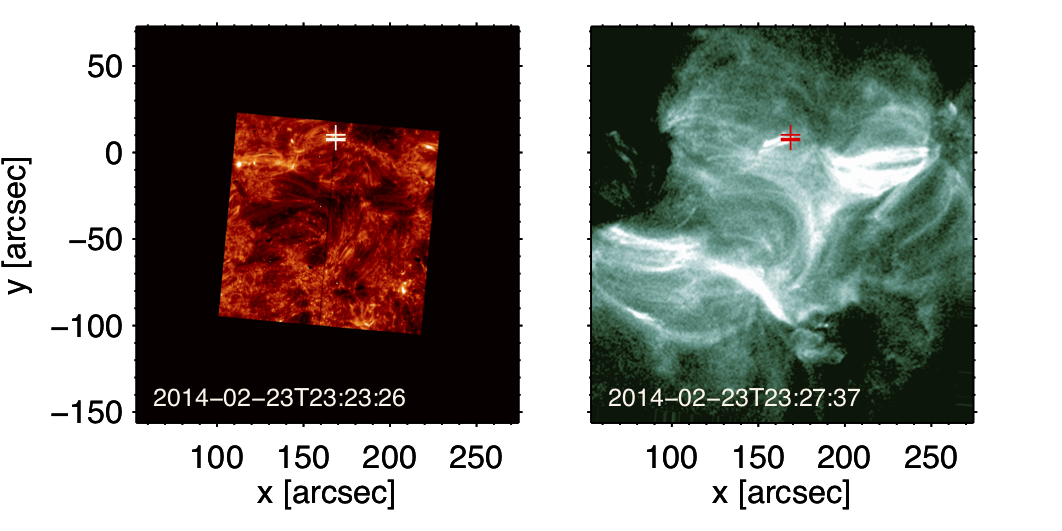}	
    \vspace{-0.3cm}
	\includegraphics[width=14cm]{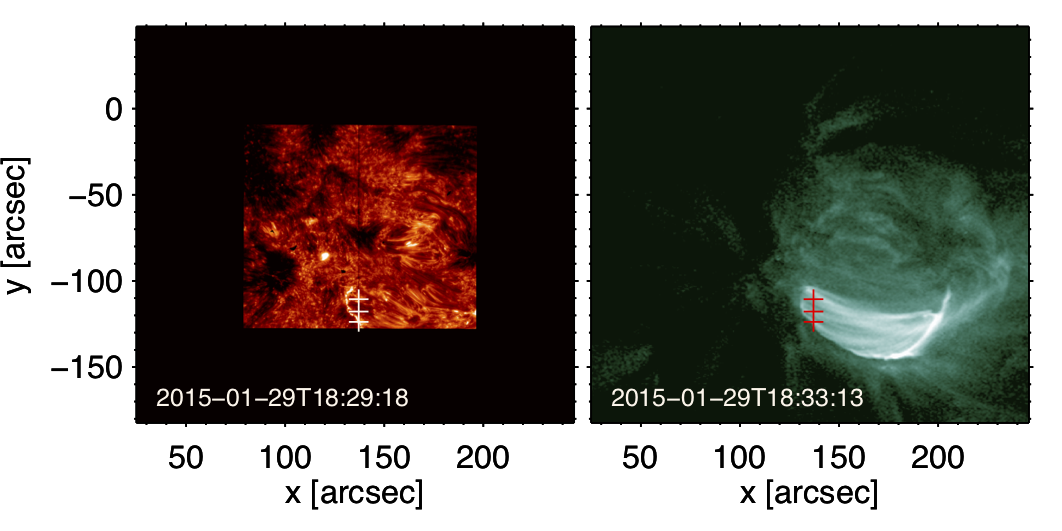}	
	\caption{\iris\ 1400\AA\ slit-jaw images ({\em left}) and \aia\ 94\AA\ images ({\em right}) for two events showing rapid moss brightenings at the loop footpoints (observed in the \siiv\ $\sim 1400$\AA\ transition region emission) followed (note the different times of the two images) by transient brightenings of the overlying hot ($\gtrsim 5-8MK$) loops (observed in the hot 131\AA\ and 94\AA--shown here-- \aia\ coronal images; see also \citealt{Reale2019a}). The images are coaligned (see text for details) and the crosses show, in both images (in white on the \iris\ images, and in red on the \aia\ 94\AA\ images, for better contrast), the location of the observed moss brightenings analyzed in details in this paper (see also Fig.\ref{fig_lc_spec1} for the event of 2014-02-23--here in the {\em top row}--, and Fig.\ref{fig_lc_spec2} for the event of 2015-01-29--here in the {\em bottom row}).}
	\label{fig_obs}
\end{figure*}

We use \iris\ calibrated level 2 data, which have been processed for dark current, flat field, and geometrical corrections \citep{DePontieu2014}, and \aia\ level 1.5 data, which have been processed for the removal of bad-pixels, despiked, flat-fielded, and image registered (coalignment among the channels, with adjustment of the roll angle and plate scales; \citealt{Lemen2012}). \aia\ datacubes in the IRIS field-of-view, already coaligned with IRIS, and in IRIS format, are now available from the IRIS data search pages \footnote{http://iris.lmsal.com/search/} (and their description and usage is discussed in detail in an \iris\ technical note \footnote{https://www.lmsal.com/iris\_science/doc?cmd=dcur\&proj\_num=IS0452\&file\_type=pdf}). The absolute wavelength calibration is already automatically applied to \iris\ level 2 spectral data, using neutral lines (such as e.g., the \oi\ 1355.6\AA\ line), which are expected to have, on average, intrinsic velocity of less than 1 km/s (when averaged along the slit; e.g., \citealt{DePontieu2014}). However, we also check this calibration by fitting additonal available neutral/photospheric lines (e.g., \feii\ 1392.8\AA\ and \si\ 1401.5\AA\ in the FUV, and \ni\ 2799.5\AA\ in the NUV).  We also apply a deconvolution method to the \iris\ spectra, to take into account the effect of the instrument Point Spread Function (PSF), as modeled by \cite{Courrier2018}. \cite{Courrier2018} used IRIS observations of a Mercury transit in 2016 to evaluate and model the on orbit IRIS PSF, and provided deconvolution routines for application to IRIS observations (iris\_sg\_deconvolve.pro). \cite{Courrier2018} noted that spectra of bright and compact brightenings, like the footpoint brightenings we focus on, might be particularly sensitive to PSF effects. We follow the suggestions of \cite{Courrier2018} to limit the number of iterations of the deconvolution algorithm to optimize the deconvolution and avoid fitting noise; in our cases we find that typically $\lesssim 10$ iterations are sufficient in both FUV and NUV.

We also analyze \aia\ coronal imaging data in six narrow extreme ultraviolet (EUV) channels (94\AA, 131\AA, 171\AA, 193\AA, 211\AA, 335\AA), which sample the transition region and corona across a broad temperature range \citep{Boerner2012,Boerner2014}, and are characterized by $\sim 0.6$~\arcsec pixels, and 12~s cadence. We also use the \aia\ 1600\AA\ images, which have a lower temporal cadence (24~s), for co-alignment between the \aia\ and \iris\ datasets, by applying a standard cross-correlation routine (tr\_get\_disp.pro, which is part of the IDL SolarSoftware package; \citealt{SSW}) to the \iris\ 1400\AA\ and the \aia\ 1600\AA\ images, which typically show many morphological similarities. The co-alignment procedure allows us to remove residual small shifts and the relative roll angle between the two instruments. For the analysis of the temporal evolution of the coronal emission we also correct the \aia\ time series for solar rotation.

\begin{figure*}
	\centering
         \hspace{-0.5cm}\includegraphics[width=14.2cm]{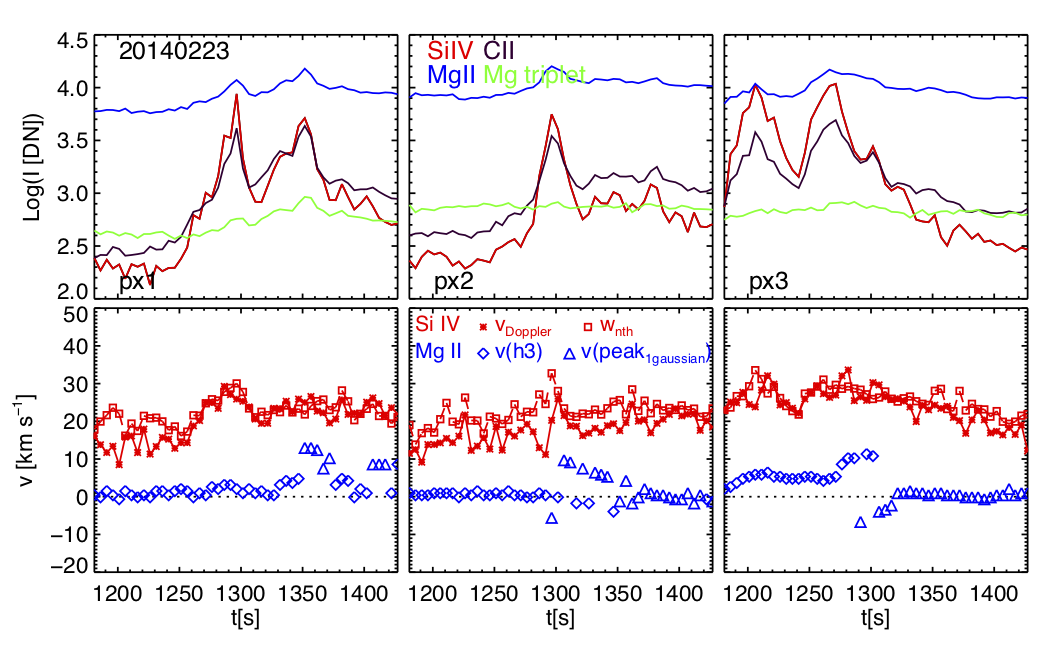}
         \includegraphics[width=15cm]{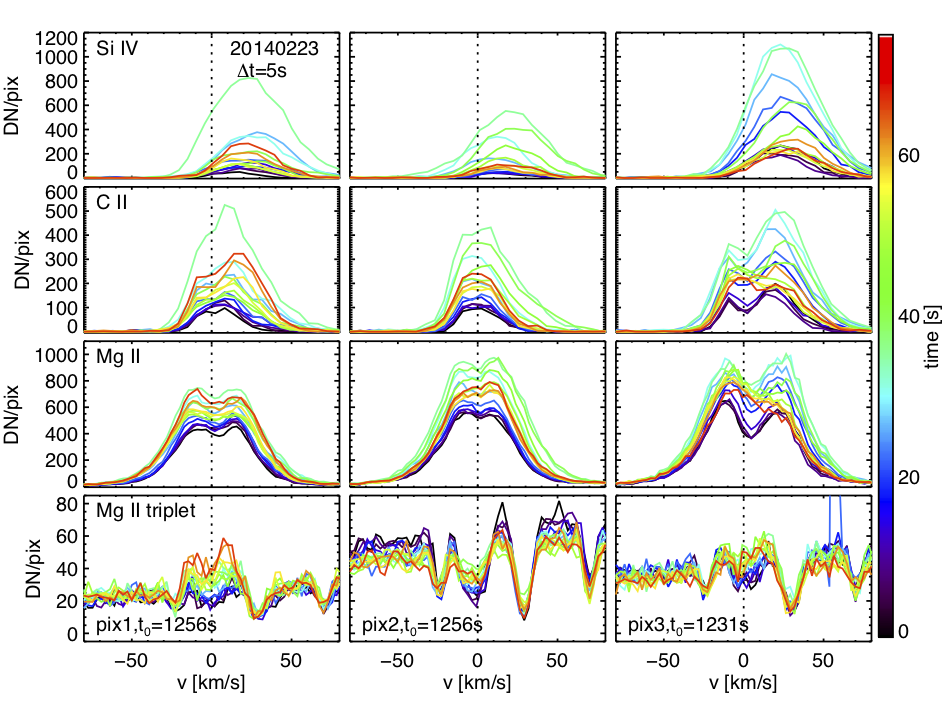}\vspace{-0.2cm}
	\caption{Temporal evolution of \iris\ spectral observables, for 3 different pixels (each corresponding to a column) during the moss brightening event observed on 2014-02-23 (as marked in the {\em top row} of Fig.~\ref{fig_obs}; event 1 of Table~\ref{table_obs}). The footpoints locations are ordered (pix1, pix2, pix3) with ascending value of solar y. {\em Top two rows:} \iris\ spectral line intensities (integrated over spectral profile; {\em top}), and \siiv\ and \mgii\ spectral moments ({\em second row}) -- Doppler velocity and non-thermal line width (\wnth) from single Gaussian fits to the \siiv\ 1402.77\AA\ spectra, and Doppler velocity of $h3$ or peak of single Gaussian for \mgii\ (see text for details). Negative values (blueshifts) of velocity correspond to upflows, and positive Doppler shifts (redshifts) to downflows. {\em Bottom four rows}: spectral line profiles, as a function of velocity, for each time step, color coded as indicated in the color bar on the right. }
	\label{fig_lc_spec1}
\end{figure*}

In the main text of this paper we describe in detail two examples (no.~1 and 6 of Table~\ref{table_obs}) and show relevant plots, while the corresponding figures for the other datasets are in Appendix~\ref{appendix:data} to improve the readability of the paper.

Fig.~\ref{fig_obs} shows TR and coronal images for these two datasets: the \iris\ slit-jaw images (SJI) in the 1400\AA\ passband are typically showing TR ($T \sim 10^5$~K) emission (including the strong \siiv\ 1402.77\AA\ line, which is one of the main targets of our analysis), while the \aia\ 94\AA\ passband has both a cool ($T \sim 10^6$~K) and a hot ($T \sim 8$~MK) peak in the temperature response \citep[e.g.,][]{Boerner2014,Cheung2015} but its emission in the bright transient loops in the active region core is generally dominated by hot plasma \citep[e.g.,][]{Testa2012b}. The TR footpoint brightenings precede the peak of the hot coronal emission, therefore we show coronal images for later times (seconds to few minutes) so that the hot coronal loops are brighter and better visible.
In this Figure (as well as in the other Figures~\ref{fig_obs2}-\ref{fig_obs10} in Appendix~\ref{appendix:data}) we mark (with crosses) the locations, at the loop footpoints, for which we discuss in detail the temporally resolved \iris\ spectra of the rapid brightenings. The moss brightenings are often observed at different times (i.e., they are not all co-temporal), and since we show a single image (both for the \iris\ 1400\AA\ SJI, and the \aia\ 94\AA), this  accounts for some discrepancy between the location of the TR brightenings and the apparent footpoints of the hot loops, which are rapidly evolving (e.g., bottom row of Fig.~\ref{fig_obs}).
 
For each event we derive a peak emission intensity in the \aia\ 94\AA\ coronal band, and an estimate of the duration of the coronal event (column 8 and 9 of Table~\ref{table_obs} respectively). We also use the \aia\ 94\AA\ images to estimate the projected loop length (in the plane-of-the-sky; column 7 of Table~\ref{table_obs}). The derived projected loop lengths should only be considered as approximate values given that several of these events involve a large number of loops, with a broad range of geometries, and it is sometimes difficult to clearly trace the loop rooted in a particular moss location. However, even with these uncertainties, the values we find show that rapid moss variability occurs at the footpoints of hot coronal loops with a wide variety of properties. 
The two examples in Fig.~\ref{fig_obs} showcase this large range of coronal properties, with event~1 being characterized by short loops, with coronal emission which is relatively faint and short-lived (few minutes), and event~6 involving instead a large number of longer loops with high and long lasting coronal emission. In fact, hot loops of event~6 appear to be bright, though highly variable, for a long time in the \aia\ 94\AA\ passband, and the heating episodes observed by \iris\ for this event appear to be part of an ensemble of repeated heating releases.

\begin{figure*}
	\centering
         \hspace{-0.5cm}\includegraphics[width=14cm]{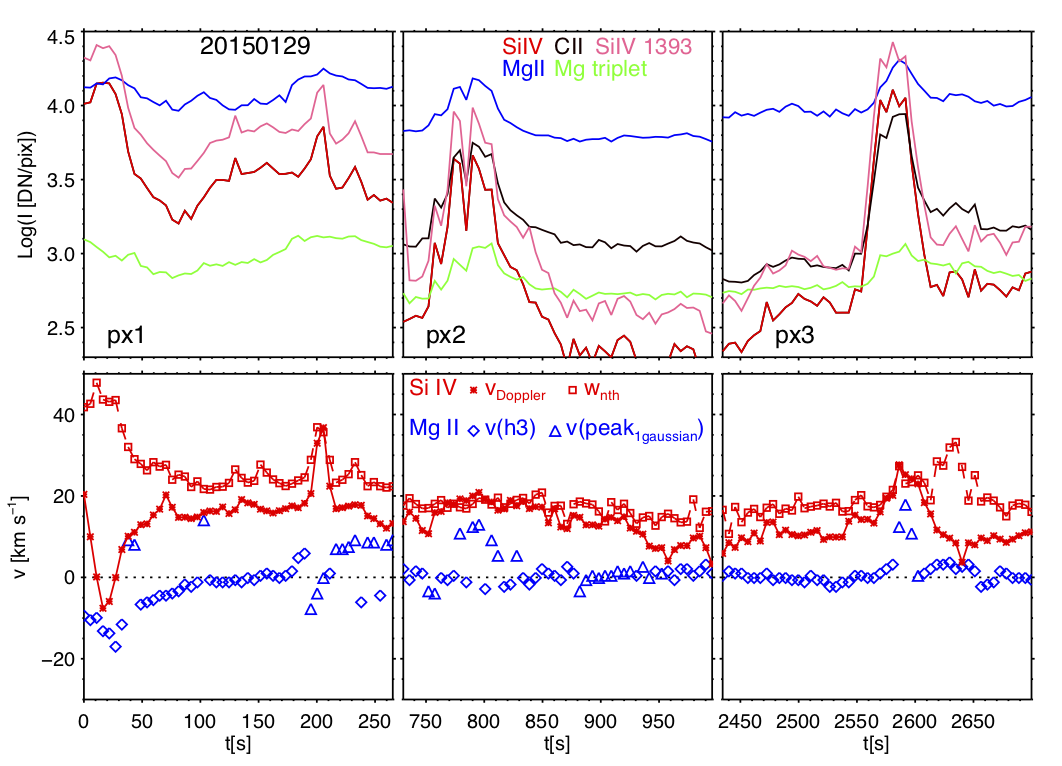}
         \includegraphics[width=15cm]{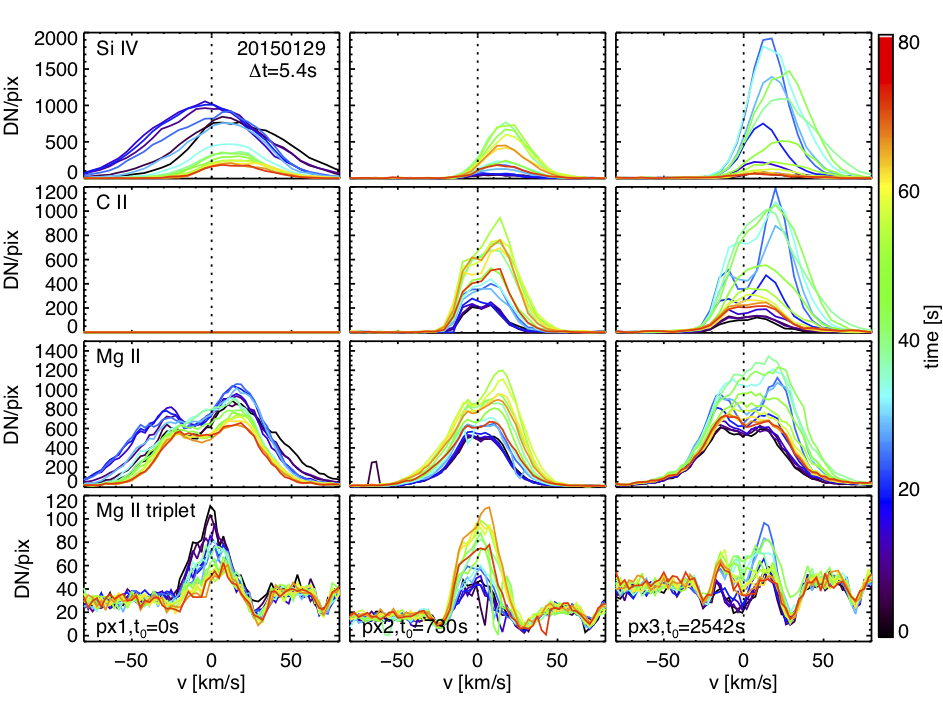}\vspace{-0.2cm}
	\caption{Temporal evolution of \iris\ spectral observables, for 3 different pixels (each corresponding to a column) during the moss brightening event observed on 2015-01-29 (as marked in the {\em bottom row} of Fig.~\ref{fig_obs}, and ordered by increasing value of solar y; ; event 6 of Table~\ref{table_obs}), in the same format as for Fig.\ref{fig_lc_spec1}. We note that for px1 the \cii\ spectrum is not available as that footpoint was at the edge of the detector and the shorter wavelength end of the FUV spectrum was outside the detector.
	}
	\label{fig_lc_spec2}
\end{figure*}

In Figures~\ref{fig_lc_spec1} and~\ref{fig_lc_spec2} we show the \iris\ spectral observations for the footpoint brightenings marked in Fig.~\ref{fig_obs}, and the results of our spectral analysis. In our investigation we particularly focus on the following lines: the \siiv\ 1402.7\AA\ (and \siiv\ 1393.8\AA, whenever included in the \iris\ OBSID line list) transition region line ($\log T[K] \sim 4.9$), and the chromospheric \cii\ 1335\AA\ ($\log T[K] \sim 4.5$; we note that \cii\ has both a chromospheric and TR nature, see e.g., \citealt{Rathore2015}) and \mgii\ h\&k (around 2803\AA\ and 2796\AA\ respectively; $\log T[K] \sim 4.3$) and \mgii\ 2798.8\AA\ triplet ($\log T[K] \sim 4.3$) lines.
We show the lightcurves obtained by integrating the line intensity over the spectral profile. For the plots of both the lightcurves and the spectral profiles we did not normalize by exposure time in order to maintain the information on signal-to-noise ratios typical of different observations.
To obtain uniform estimates of the spectral properties of the \siiv\ and \mgii\ lines for the whole sample, we carried out spectral fits as follows. For the \siiv\ lines we perform a single Gaussian fit, and derive the Doppler shift ($v_{\rm Doppler}$) of the line centroid and the line width. From the measured line width we derive the non-thermal line broadening (\wnth), i.e., the broadening in excess of the thermal ($w_{\rm th} = \sqrt{2 K_{\rm B} T/m_{\rm ion}}$), and instrumental (for \iris\ is of the order of 3.5~km~s$^{-1}$; \citealt{DePontieu2014}) broadening. Note that here we refer to line broadening as $w_{\rm 1/e}$ (in km s$^{-1}$), which is the $1/e$ spectral line width, and which corresponds to $\sqrt{2} \times$ the Gaussian $\sigma$, and FWHM/$(2 \sqrt{ln(2)})$.
The \mgii\ lines have more complex spectral profiles, and different \iris\ routines are available in the \iris\ branch of SolarSoft for the \mgii\ spectral fitting using different approaches. Here we use the routine iris\_master\_fit, described in \cite{Schmit2015}, which has been optimized for the fitting of observed \iris\ \mgii\ spectral profiles. The fit typically uses a double Gaussian plus a linear function, however, in cases where the observed profiles is single peaked it provides the parameters of a single Gaussian fit. Therefore, in our plots of Fig.~\ref{fig_lc_spec1} and ~\ref{fig_lc_spec2} (and Fig.~\ref{fig_obs2}-\ref{fig_obs10} in Appendix~\ref{appendix:data}) for \mgii\ we show the Doppler shift of the \mgii\ $h3$ feature (${\rm v(h3)}$) or the Doppler shift of the centroid of the single peak (${\rm v (peak_{\rm 1gaussian})}$) , depending on whether the profile is double or single peaked respectively.
While these spectral fits generally provide a good estimate of the spectral moments of the \siiv\ and \mgii\ profiles, some particularly unusual and complex profiles are not well fitted by these fitting functions.

\begin{figure}
	\centering
	\begin{minipage}{9cm}
	\centering
         \includegraphics[width=9cm]{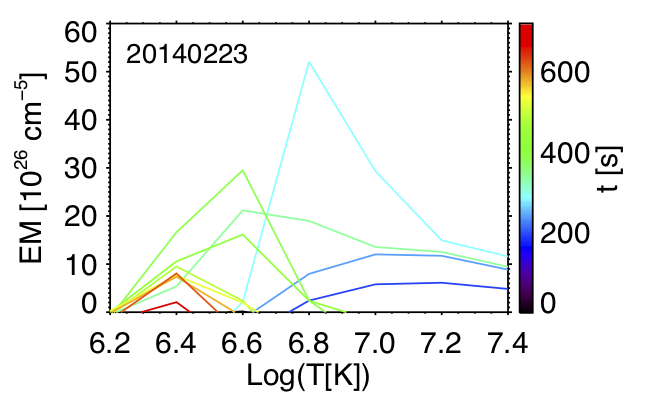}
         \end{minipage}	
	\hspace{-0.5cm}\begin{minipage}{9cm}
	\centering
         \includegraphics[width=9cm]{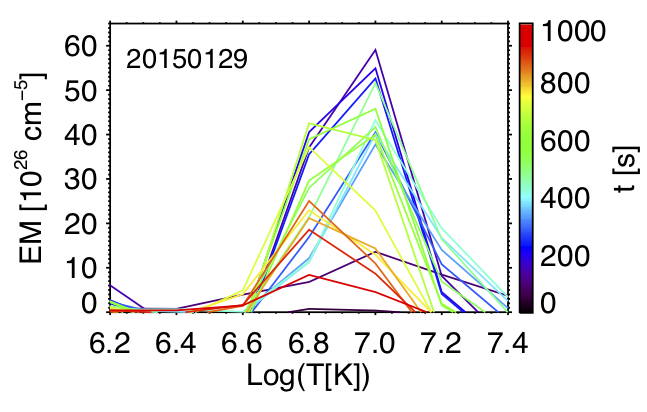}
        	\end{minipage}	
	\caption{Emission measure (EM) distributions vs temperature (after subtraction of the EM prior to the heating event) at progressive times, color coded as indicated by the color bars on the right, for the two events of Fig.~\ref{fig_obs}. The reference times are  2014-02-23T23:21:43, and 2015-01-29T18:28:07 respectively.}
	\label{fig_dem}
\end{figure}

The observed chromospheric and TR footpoint brightenings are typically characterized by an increase in \siiv\ intensity of an order of magnitude, though this increase factor ranges from less than 2 in the smallest events (e.g., event 5 in Fig.~\ref{fig_obs5}, for which however the brightest footpoints are missed by the \iris\ slit), to about 2 orders of magnitude (or more, for e.g., some examples shown in \citealt{Testa2014}) in the brightest events (e.g., event 10 in Fig.~\ref{fig_obs10}).  Also, the duration of the brightenings is generally short ($\lesssim 60$s), and very similar for all of the events.
Generally the \siiv\ brightenings are accompanied by significant brightenings in the \mgii\ and \cii\ chromospheric lines as well, however in our sample there are also two events for which that is not the case and either \mgii\ (event 7; Fig.~\ref{fig_obs7}) or \cii\ (event 5; Fig.~\ref{fig_obs5}) present only a very small increase in intensity, if any.
The \mgii\ triplet (around 2799\AA) which, as we we will discuss in more detail in the following sections, provides interesting diagnostics of low chromospheric heating \citep{Pereira2015}, is observed in emission for some locations in a subset of the events in our sample (4 out of 10 events).
As in the event we studied in detail in \citep{Testa2014}, the brightenings for which we have \iris\ spectra, show a broad range of \siiv\ Doppler shifts, relative to the line shift before the event, including redshifts (i.e., downflows), blueshifts (i.e., upflows), or no significant shifts. Also, the \siiv\ lines during most of the brightenings show small or no significant increase in broadening, however with a few notable exceptions including for event 6, for which a footpoint brightening shows very broad but still fairly (single) Gaussian profiles (pix1, Fig.~\ref{fig_lc_spec2}), and event 8 (Fig.~\ref{fig_obs8}) and 9 (e.g., pix3, Fig.~\ref{fig_obs9}), for which the \siiv\ profiles during the event become quite complex with several components.

\begin{deluxetable*}{ccccccc}
\tablecaption{RADYN simulations discussed in detail.}
\tablehead{
\colhead{Model\tablenotemark{a}} &  \colhead{Name} &  \colhead{E$_T$} & \colhead{F} & \colhead{E$_C$} &  \colhead{$\delta$} & \colhead{duration}   \\
 &   &  \colhead{[10$^{24}$ erg]} &  \colhead{[erg s$^{-1}$ cm$^{-2}$]} &  \colhead{[keV]} & & \colhead{[s]} 
}
\startdata
C1  &  TC            & 6  & - & - & -&10\\
E1  &  5keV          & 6  & 1.2~$\cdot$~10$^{9}$ & 5 & 7& 10\\
E2  &  10keV         & 6  & 1.2~$\cdot$~10$^{9}$ & 10 & 7& 10 \\
E3  &  15keV         & 6  & 1.2~$\cdot$~10$^{9}$ & 15 & 7& 10 \\
H1  &  10keV+TC      & 6  & 0.6~$\cdot$~10$^{9}$ & 10 & 7& 10 \\
E4  &  10keV, 30s    & 6  & 0.4~$\cdot$~10$^{9}$ & 10 & 7& 30 \\
H2  &  10keV+TC, 30s & 18 & 0.6~$\cdot$~10$^{9}$ & 10 & 7& 30 \\
E5  &  15keV, 30s    & 18 & 1.2~$\cdot$~10$^{9}$ & 15 & 7& 30 \\
E6  &  10keV, 20s\_p & 6  & 0.6~$\cdot$~10$^{9}$ & 10 & 7& 20 + 60\tablenotemark{b} \\
\enddata
\label{table_sims}
\tablenotetext{a}{Model labels are: C for heating by thermal conduction only, E for heating by accelerated electrons only, H for hybrid models with a mix of heating by conduction and non-thermal particles.}
\tablenotetext{b}{Two repeated heating pulses with duration of 20~s and 60~s pause in between pulses.}
\end{deluxetable*}

\begin{deluxetable*}{lccccc}
\tablecaption{IRIS spectral properties \tablenotemark{a} of RADYN simulations of Table~\ref{table_sims}.}
\tablehead{
\colhead{Model} & \multicolumn{2}{c}{\siiv} & \colhead{\cii\  \tablenotemark{b}} & \colhead{\mgii\  \tablenotemark{b}} &  \colhead{\mgii\ triplet}   \\ \cline{2-3}
 &  \colhead{Doppler} & \colhead{line shape} & & & \vspace{-0.3cm} \\
 &  \colhead{shift} &  & & & 
}
\startdata
C1  &  red  & & weak; redshift; stronger rp & weak; peculiar shape; stronger bp & \\
E1  &  red & multi-component &  broad; redshift; stronger rp & broad; redshift;  blue tail & \\
E2  &  blue &  & symmetric & stronger bp & Y \\
E3  &  blue & weak \siiv\ & stronger rp & symmetric & Y \\
H1  &  blue & multi-component & slightly stronger rp & slightly stronger bp &  \\
E4  &  blue & blue tail & stronger rp  & symmetric & \\
H2  &  blue & multi-component & $\sim$ symmetric; red tail & broad; redshift; blue tail & Y \\
E5  &  blue & & small blueshift; red tail & complex profiles; strong broad bp & Y  \\ 
\multicolumn{1}{l|}{E6 \tablenotemark{c}}  &  blue & blue tail & stronger rp & symmetric; small redshift &  \\
\multicolumn{1}{c|}{}    &  blue  & with red tail & small redshift; broad, red tail & symmetric & \\
\enddata
\label{table_sims_iris}
\tablenotetext{a}{The values refer to the spectra at peak emission.}
\tablenotetext{b}{"bp" and "rp" are respectively blue and red peak of these chromospheric lines.}
\tablenotetext{c}{For this model the two heating pulses are characterized by quite different properties, so we describe each in a row.}
\end{deluxetable*}

In \cite{Reale2019a} we have analyzed in detail the coronal emission in the overlying loops, and its evolution, and speculated on the possible role of large-angle magnetic reconnection for these events. Therefore here we only briefly derive and discuss the coronal properties of these coronal heating events showing significant footpoint brightenings. We use the \aia\ timeseries (in the 94\AA, 131\AA, 171\AA, 193\AA, 211\AA, 335\AA\ passbands) to derive the emission measure as a function of temperature, using the inversion method of \cite{Cheung2015}.  The observed emission ($I_{i}$, in units of ${\rm DN~s^{-1}~pix^{-1}}$) in the \aia\ narrow-band EUV channels depends on the thermal properties of the optically thin coronal plasma in the pixel, as $I_{i} = \int_{T} R_{i}(T)~DEM(T)~\,dT$, where $R_{i}(T)$ is the response function in a given passband (in units of ${\rm DN~cm^{5}~s^{-1}~pix^{-1}}$), and the differential emission measure (in units of ${\rm cm}^{-5}~{\rm K}^{-1}$) is defined by $DEM(T)~\,dT = \int_{z} n_e^2(T)~\,dz$, where $n_e^2(T)$ is the electron density of the plasma at temperature T. The distribution of emission measure (EM, in units of ${\rm cm}^{-5}$) as a function of temperature which we will show in several plots throughout this paper is obtained by integrating the $DEM(T)$ over 0.2 $\log T$ temperature ranges.
In Figure~\ref{fig_dem} we show the temporal evolution of the emission measure distribution for the coronal loops of the events shown in Fig.~\ref{fig_obs}-\ref{fig_lc_spec2}. These plots show coronal temperatures of $\log T \sim 6.8-7.0$ associated with these heating events, as also discussed in more detail by \cite{Reale2019a}.

\section{RADYN numerical simulations}
\label{sims}

In our previous work \citep{Testa2014,Polito2018} we demonstrated the importance of numerical simulations for the interpretation of the observations. In \cite{Polito2018} we used RADYN 1D HD loops simulations to investigate the plasma response to impulsive heating, for varying heating properties and initial conditions. In particular we focused on the predicted optically thin emission (TR emission in the \iris\ \siiv\ line and coronal emission in the \aia\ 94\AA\ passband), and we also discussed some properties of the predicted \iris\ \mgii\ emission. 

In the previous section we have presented an analysis of the \iris\ chromospheric and TR spectral observations, which can provide much tighter constraints on the properties of the heating. Therefore we expanded on our previous investigation of RADYN nanoflare heated loops, by calculating the synthetic chromospheric emission in \cii\ and \mgii\ triplet for the simulations previously discussed in \cite{Polito2018}. 
The very broad variety of properties of the spectral observations analyzed in the previous section (section~\ref{observations}) motivated us to also explore a few additional simulations to explore different properties of the heating including: (a) combining heating by thermal conduction and non-thermal particles; (b) longer heating duration of 20-30~s; (c) repeated heating in the same loop (in particular, two 20s heating episodes separated by 60~s intervals when the heating is switched off).

\begin{figure*}
	\centering
	\vspace{-0.8cm}
	\includegraphics[height=18.3cm]{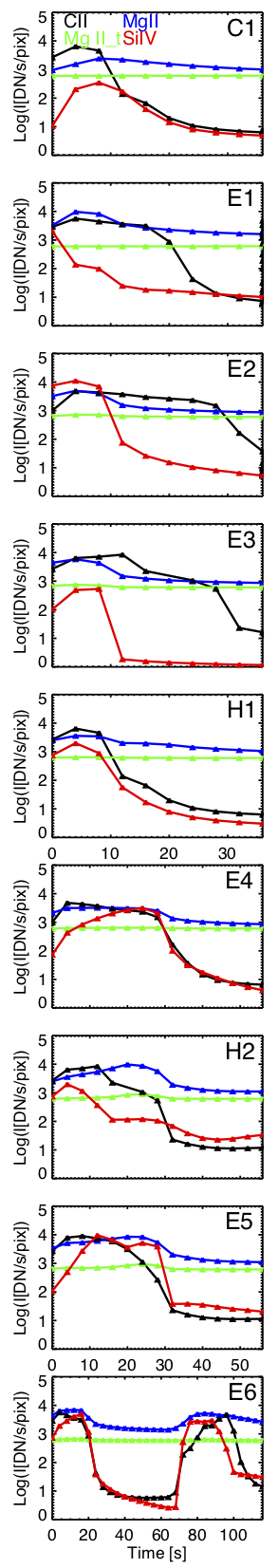} 
	\includegraphics[height=18.5cm]{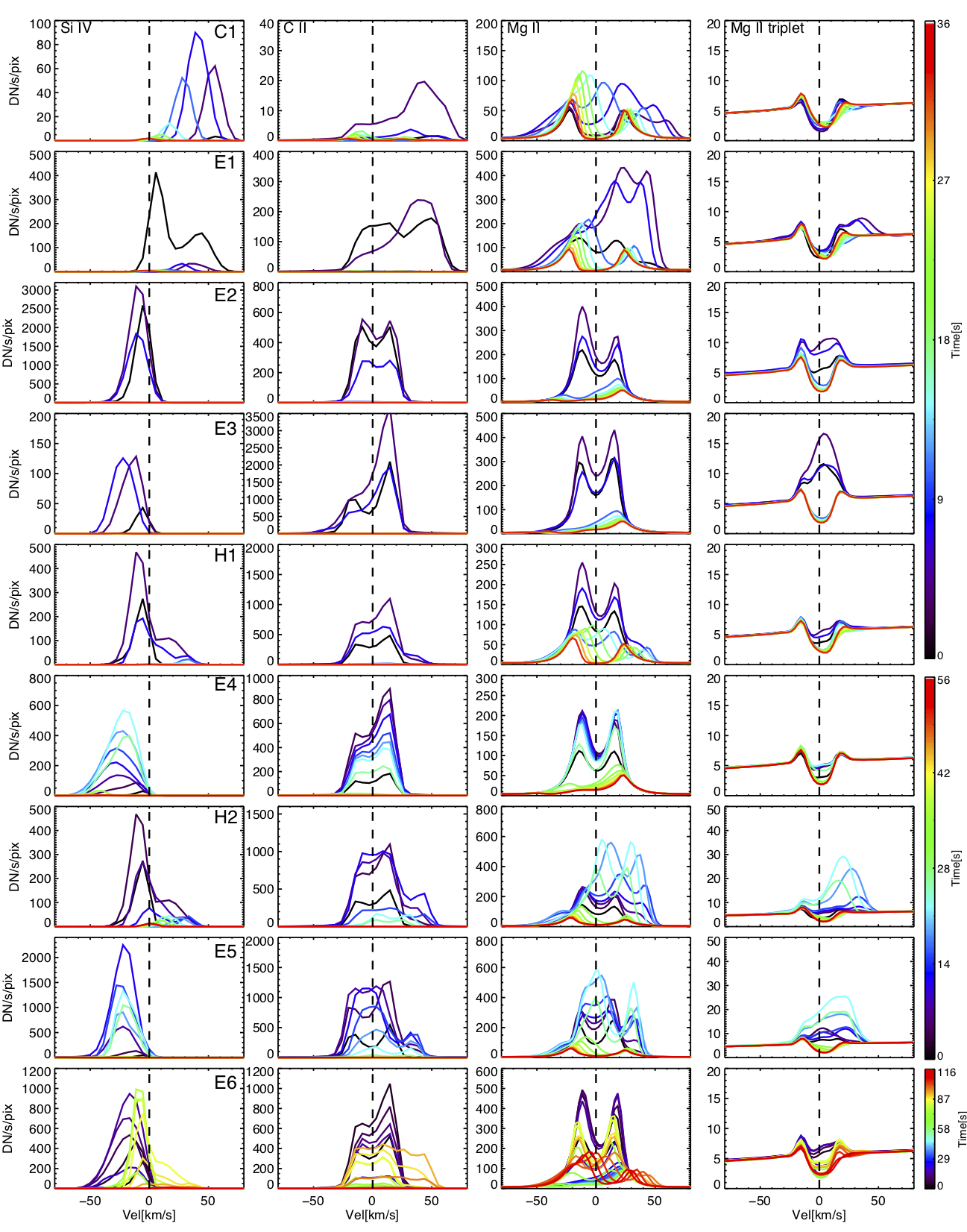}
	\caption{{\em Left column}: Synthetic lightcurves in \iris\ spectral lines as derived from the RADYN simulations of Table~\ref{table_sims}. These predictions are calculated by assuming an integration time (and cadence) of 4~s (i.e., the data point at t=0 corresponds to the average intensity in the first 4~s since the onset of heating), and can be directly compared with the observed lightcurves (top panels of Figures~\ref{fig_lc_spec1} and~\ref{fig_lc_spec2}, and Figures~\ref{fig_obs2}-\ref{fig_obs10} of Appendix~\ref{appendix:data}).
	{\em Right four columns:} Temporal evolution of corresponding synthetic \iris\ spectra, in the same format as the observed spectra (bottom plots of Figures~\ref{fig_lc_spec1} and~\ref{fig_lc_spec2}, and Figures~\ref{fig_obs2}-\ref{fig_obs10} of Appendix~\ref{appendix:data} for the other studied events). These synthetic spectra are obtained by assuming an integration time (as well as a cadence) of 4~s, which is similar to most of the observations studied here (see Table~\ref{table_obs}). We note that, due to the different duration of the heating in the different models the timescales are different for the first 5 rows (36~s; models C1, E1, E2, E3), row 6-8 (56~s; models E4, H2, E5), and the last row (116~s, model E6).
	}
	\label{fig_radyn_1}
\end{figure*}

The details of the RADYN numerical code \citep{Carlsson1997,Allred2005,Allred2015} and the general description of how these simulations of nanoflare heated loops are run are discussed in detail in \cite{Polito2018}, so here we only provide a brief description. 
The RADYN code solves the 1D equation of charge conservation and the level population rate equations for the magnetically confined plasma; the loop's atmosphere encompasses the photosphere, chromosphere, TR, and corona. RADYN is ideal for our investigation because of (a) it includes non-local thermodynamic equilibrium (non-LTE) radiative transfer, which is necessary to model the chromospheric emission, and (b) it allows to model heating by non-thermal electron (NTE) beams \citep{Allred2015}. As discussed by \cite{Polito2018} we added chromospheric heating to obtain a more realistic plage-like atmosphere as in \cite{Carlsson2015}.
The \iris\ \siiv\ optically thin emission is synthesized from the model by using atomic data from CHIANTI \citep{chianti,chianti7,chianti8} and assuming ionization equilibrium, as in \cite{Polito2018}. The \iris\ optically thick \cii, \mgii, and \mgii\ triplet emission is synthesized by using RH1.5, which is a massively-parallel code for polarised multi-level radiative transfer with partial frequency distribution \citep{PereiraUitenbroek2015}.
We have performed dozens of RADYN simulations of nanoflare-heated loops investigating a broad parameter space, including different:
\begin{itemize}
\item nanoflare energies: 10$^{24}$ to 10$^{26}$ ergs
\item loop top temperatures : T$_{\rm LT}$ = 1, 3, and 5~MK 
\item half-loop lengths L/2 = 15, 50 and 100~Mm
\item heating models: (a) thermal conduction (TC), (b) electron beam (EB) heating with different energy cut-offs E$_C$ = 5, 10 and 15 keV, and (c) TC + EB hybrid models
\end{itemize}

Some of these simulations have already been presented in detail in \cite{Polito2018}.
In the following we discuss in detail nine of these (dozens of) simulations summarized above, for which the synthetic spectra have characteristics similar to the observed spectra in terms of e.g., duration of the brightenings and line intensities (see discussions in Sec.~\ref{discussion} and \ref{conclusions}). In particular, as summarized in Table~\ref{table_sims}, we selected four of the simulations presented in \cite{Polito2018} -- with an initially empty cool loop ($T \sim 1$~MK), of semi-length of 15~Mm, heated for 10~s by TC or NTE with 5/10/15~keV energy cutoff --, and five new simulations, all run on the same cool initial condition: (1) 10~s heating with TC and NTE with 10~keV cutoff (and total energy of $6 \times 10^{24}$~ergs, as in the previous runs), (2) intermittent heating (two 20~s heating episodes with 60~s pause in between) by NTE with 10~keV cutoff, (3) 30~s heating by NTE with 10~keV cutoff, (4)  30~s heating with TC and NTE with 10~keV cutoff, and total energy of $1.8 \times 10^{25}$~ergs, (5) 30~s heating by NTE with 15~keV cutoff, and total energy of $1.8 \times 10^{25}$~ergs.

Figure~\ref{fig_radyn_1} shows the predicted \iris\ lightcurves (left column) for the nine selected RADYN simulations, synthesized by assuming an integration time (and cadence) of 4~s. A qualitative comparison of these synthetic lightcurves with the observed ones (shown in Fig.~\ref{fig_lc_spec1}, \ref{fig_lc_spec2}, \ref{fig_obs2}-\ref{fig_obs10}) shows in general a very good agreement between simulations and observations in terms of intensities and durations of the brightenings. As observed, the relative increase of emission in \mgii\ is, typically, significantly smaller than in \siiv\ or \cii.
We will discuss this comparison in more detail in section~\ref{discussion}.

Figure~\ref{fig_radyn_1} shows the temporal evolution of the synthetic \iris\ spectra for the selected simulations, in the same format used for the observations in Fig.~\ref{fig_lc_spec1} and~\ref{fig_lc_spec2}. The spectra are obtained by assuming an integration time and cadence of 4~s, similar to many of the studied observations, and applying the appropriate thermal and \iris\ instrumental broadening, and assuming a spectral bin of 0.025\AA\ for all spectral lines (i.e., corresponding to a binning of $\times 2$ in FUV and original spectral resolution in NUV, which is a quite typical observing mode for these nanoflare observations, due to the typically higher S/N in NUV).

First of all, the synthetic spectra show a very broad variety of spectral line profiles, similarly to the observations. For instance, the \siiv\ emission in several cases shows a complex transient multiple component profile (e.g., 5~keV, 10~keV+TC).

The \mgii\ emission sometimes shows unusual profiles, similar to what is occasionally observed (see e.g., Fig.~\ref{fig_obs5}, ~\ref{fig_obs9}). We note that, despite our efforts in creating a more realistic plage-like atmosphere, some of the observed filled in (single Gaussian) \mgii\ profiles are not easily reproduced by the simulations (see also discussion in \citealt{Polito2018}). 

The \mgii\ and \cii\ synthetic profiles with a strong increase of the red peak to blue peak ratio (e.g., \cii\ profile for the 15~keV case of Fig.~\ref{fig_radyn_1}), are usually due to an upflow which causes a blueshift of the $\tau_\nu =1$ layer which causes in turn the increased absorption of the blue peak. Analogously, profiles with a higher blue peak (e.g., \mgii\ profile for the 10~keV case of Fig.~\ref{fig_radyn_1}) can be caused by downflows causing an increased absorption of the red peak.

A notable results concerns the \mgii\ triplet, which in our simulations are predicted to be strongly in emission only if high energy NTE are present. This is due to the fact that, in our modeled atmospheres, the \mgii\ triplet is much more sensitive to heating in the lower chromosphere, close to the temperature minimum, than the \cii\ and \mgii\ chromospheric lines, and it is therefore expected to be in emission only if those lower chromospheric layers are significantly heated \citep[see][for a detailed discussion of the \mgii\ triplet formation mechanism]{Pereira2015}. 
This is a very important finding as it results in a new diagnostic of the presence of NTE in these events.

In Table~\ref{table_sims_iris} we summarize some of the main spectral properties of the \iris\ synthetic spectra from our simulations, for easier comparison with the observations (see Sec.~\ref{discussion} and Table~\ref{table_obs_mod}).

\section{Discussion}
\label{discussion}

\begin{figure}
	\centering
	\includegraphics[width=6.5cm]{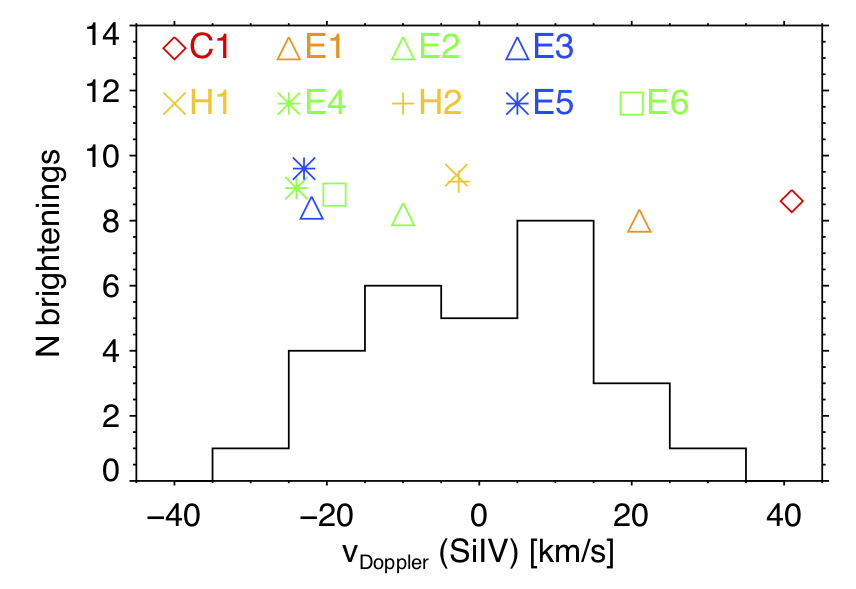} 
	\vspace{-0.2cm}
	\includegraphics[width=6.5cm]{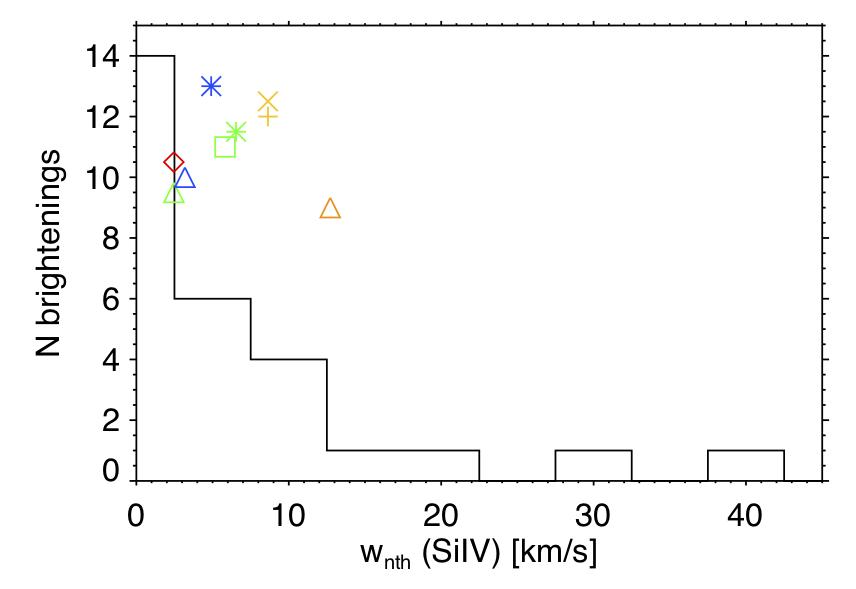}
	\vspace{-0.2cm}
	\includegraphics[width=6.5cm]{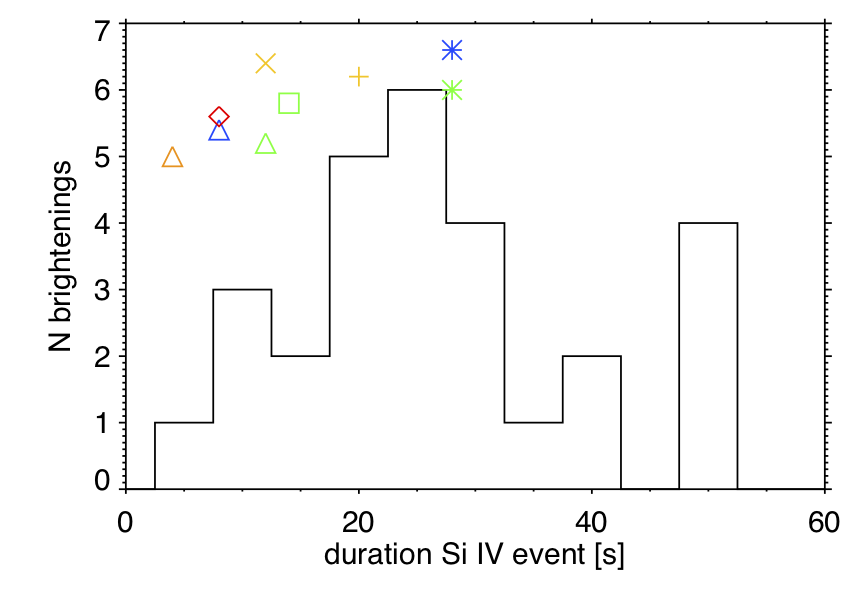}
	\vspace{-0.2cm}
	\includegraphics[width=6.5cm]{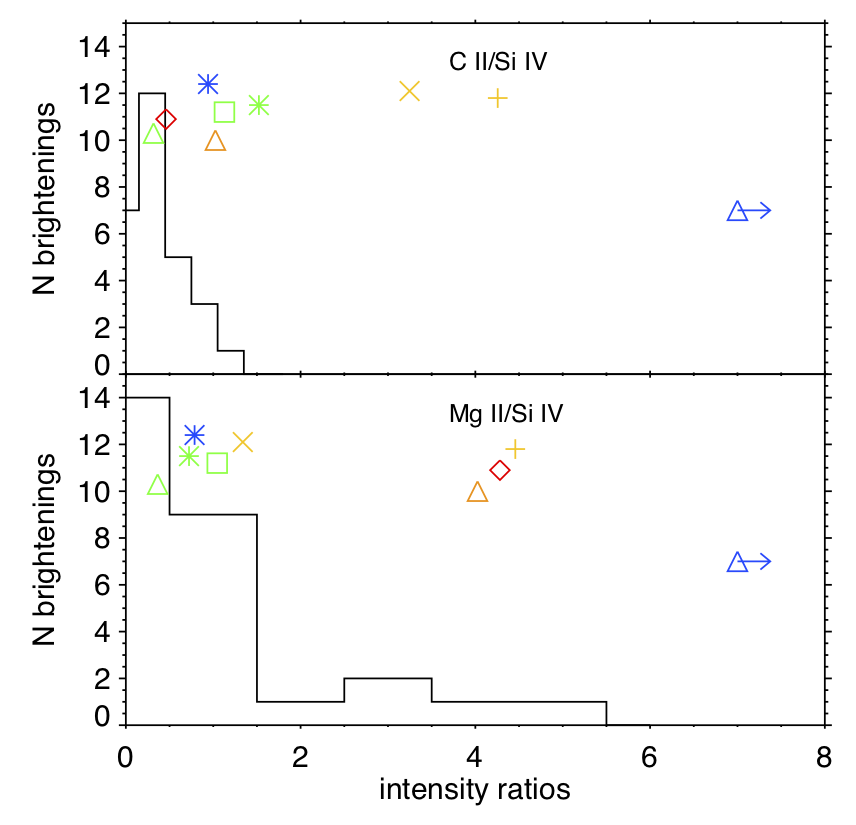}
	\caption{Histograms of observed properties of footpoints brightenings observed by \iris, and associated with coronal heating events (from {\em top}): \siiv\ Doppler velocity, \siiv\ non-thermal line broadening, duration of the moss brightening in \siiv\ (where $I > I_{\rm max}/e$), ratio of intensity maxima for \cii/\siiv\ and \mgii/\siiv. The colored symbols show the corresponding predictions of the different RADYN models (see legend in the top panel) described in the text (and Table~\ref{table_sims}-\ref{table_sims_iris}, Figure~\ref{fig_radyn_1}. Each symbol corresponds to one RADYN simulation, and we offset the values in the y direction in order to make them distinguishable even when different simulations have similar values of the shown parameters.}
	\label{fig_hist}
\end{figure}

In Sections~\ref{observations} and \ref{sims} we have discussed respectively the observed and modeled \iris\ spectral properties of transition region brightenings arising in response to impulsive coronal heating. The simulations presented in Section~\ref{sims} (and also in \citealt{Polito2018}) have not been run to forward fit the specific observations presented here, but they do explore a relatively broad parameter space and can be used for a qualitative comparison with the observations, and their interpretation. In order to investigate the qualitative agreement between the predicted and observed spectral properties we first construct histograms of the spectral properties derived from the \iris\ observations. In Figure~\ref{fig_hist} we show histograms for the Doppler velocity and non-thermal broadening (at peak intensity) of the \siiv\ line, for the duration of the TR brightening (defined as the time interval when $I > I_{\rm max}/e$ for \siiv), and for the ratios of the peak intensity values of \siiv, \cii, \mgii\ (in particular we show the \cii/\siiv, and \mgii/\siiv\ ratios), all measured from the observed IRIS spectra.
We note that here we take into account all the different footpoint brightenings, as for many events there is more than one footpoint brightening observed by \iris.
We show the \cii/\siiv, and \mgii/\siiv\ intensity ratios because the are a useful diagnostics of the relative amount of energy deposited in different atmospheric layers, as predicted from the models (see Fig.~\ref{fig_radyn_1} and Table~\ref{table_sims_iris}).
These histograms show that the observed brightenings are characterized by:
\begin{itemize}
\item \siiv\ relative Doppler velocity with a broad distribution, quite symmetric around zero, and with absolute values $|{\rm v_{Doppler}| \lesssim 40}$~km~s$^{-1}$
\item generally modest non-thermal broadening increases, with only a few footpoint brightenings (e.g., event 6, Fig.~\ref{fig_lc_spec2}) showing large values of \wnth, up to 40~km~s$^{-1}$; we note that some of these broader spectra still have Gaussian profiles (e.g., event 6, Fig.~\ref{fig_lc_spec2}), while others clearly show multiple components or more complex non-Gaussian profiles (e.g., event 8, Fig.~\ref{fig_obs8}, and event 9, Fig.~\ref{fig_obs9})
\item short duration ($\lesssim 30$~s)
\item similar intensity increases in \siiv\ and \cii\ (with \cii/\siiv\ typically $\sim 0.5$-$1$), whereas the brightening intensity in \mgii\ appears less closely tied to the \siiv\ emission
\end{itemize}
We note that we made specific choices in the definition of the variables shown in Fig.~\ref{fig_hist}, and that a single value might not adequately capture the complexity of some footpoint brightenings. For instance, for some brightenings (e.g., pix~2 and pix~3 of event 1, second row of Fig.~\ref{fig_lc_spec1}) the \siiv\ line appears to undergo an initial blueshift followed by a redshift (also reproduced e.g., by model H2; see below for further discussion).

We derive the analogous spectral properties from the synthetic spectra of the simulations (sec.~\ref{sims}), and superimpose them on the histograms of measured values in Figure~\ref{fig_hist}. Similarly to the observations we note that a single value might not capture the complexity and evolution of the predicted \iris\ spectral properties. This comparison shows that the RADYN simulations of nanoflare heated loops can reproduce most of the observed properties. 
In particular, the simulations reproduce well the observed range of: (a) duration and Doppler shifts of the \siiv\ TR emission, (b) relative ratio of chromospheric to TR emission, and (c) variability of emission. 

A closer look to this overall comparison of the observed range of parameters with the predictions of our models reveals some of the constraints the observations put on the simulations, and it also provides useful guidance on how to further extend the exploration of the parameter space with future simulations. For instance, the large redshift predicted by our TC simulation (model C1) is beyond the range of observed Doppler shifts, indicating that in these nanoflares, if/when TC (and/or NTE with low $E_C$) is dominating the energy transport, the initial density might likely be larger than assumed in these models (see e.g., for comparison the hot/dense models in \citealt{Polito2018}) and/or the energy released in the single event (and energy flux) might be smaller (see \citealt{Polito2018}).
We also note that only two of the selected models predict \siiv\ redshifts (C1, and E1), while several observations show redshifted \siiv. However, different combinations of heating -- e.g., NTE distributions, different duration of TC and NTE heating, total energy (see e.g., Fig.~15 of \citealt{Polito2018} which shows that larger events can have redshifts at larger $E_C$)-- and initial conditions can reproduce a broader range of Si IV Doppler shifts.
Similarly, the hybrid models (H1, H2) predict a larger than observed \cii/\siiv\ ratio, suggesting that more realistic models might need for instance slightly different partition of energy between NTE and TC (in the models for this paper we assume the energy is equally divided between the two transport mechanisms).
Therefore, as a follow-up work we will explore simulations using initial conditions with intermediate density/temperature between the cool/empty (1~MK) and the hot/dense (3~MK) initial atmosphere used in \cite{Testa2014,Polito2018} and here, and explore different heating properties. Additional effects, such as e.g., non-equilibrium ionization, are also expected to increase the predicted \siiv\ emission and likely bring the \cii/\siiv\ and \mgii/\siiv\ intensity ratios expected from our simulations in better agreement with the observations.

We also note that our loop models are generally characterized by modest non-thermal broadening, therefore not reproducing the large non-thermal broadening observed for a few footpoint brightenings. This discrepancy might either be due to physical processes missing from our simulations (e.g., turbulence; we note that lack of non-thermal broadening has been found also for 3D MHD simulations, as e.g., discussed by \citealt{Testa2016}) and/or imply a more complex scenario of multiple loops (possibly with different initial atmosphere) simultaneously heated with different heating properties (see e.g., \citealt{Polito2019} for an example of multi-loop 1D RADYN modeling though applied to larger flares), or, depending on the viewing angle, the presence of waves or turbulence that leads to broadening in the direction perpendicular to the field. Another limitation of our models is that the line profiles of the \mgii\ and \cii\ chromospheric emission are not always well reproduced, especially the single peaked profiles (typical of plage) as previously remarked by \cite{Polito2018} (see discussions therein). 
Finally we also note that the simulations are generally characterized by pre-nanoflare \siiv\ Doppler shift of $\sim 0$~km/s (and in Fig.~\ref{fig_hist} the observed \siiv\ Doppler shifts are the values relative to the pre-nanoflare "rest" velocity), while the observations (e.g., Figures~\ref{fig_lc_spec1} and~\ref{fig_lc_spec2}) often show significant TR Doppler shifts in quiescent conditions (see e.g., \citealt{Testa2016} and reference therein for examples of recent studies of TR Doppler shifts).

Notwithstanding their limitations, the simulations can be used as a guide to infer the properties of the heating on a case-by-case basis. However we note that here we will carry out a qualitative comparison, given that the simulations explore a limited range of parameters (e.g., initial conditions, total energy, duration of heating, mix of TC and NTE, properties of NTE,...) and therefore finding a perfect match between observations and simulations is beyond the scope of this paper. 
Here we discuss in detail the specific observational properties of the IRIS spectra for all the studied events, their similarities with the predictions from our models, and the resulting diagnostics. We summarize these results in Table~\ref{table_obs_mod}.
We categorize the events in three groups characterized by: (a) strong evidence of NTEs, (b) likely presence of NTEs, and (c) lack of significant evidence of NTEs:

\begin{itemize}
\item {\bf Strongest evidence of NTEs (both \mgii\ triplet emission and \siiv\ blueshifts):}
    \begin{itemize}
        \item {\bf Event 1} (Fig.~\ref{fig_obs}, top, and Fig.~\ref{fig_lc_spec1}) is a short lived heating event in small loops, but it nevertheless shows significant \mgii\ triplet emission at (at least) one of the footpoints, which, as discussed in \S\ref{sims} points to the presence of NTE. The chromospheric emission in px~1 is similar to models E2-E3, though the lack of corresponding \siiv\ blueshift suggests a mix with heating by TC (e.g., H1). The other two footpoints (px~2 and px~3) show a peculiar initial relative blueshift in \siiv\ immediately followed by a redshift similar to model H2 (though we observe \siiv\ profiles generally well approximated by a Gaussian). For these footpoints also the chromospheric emission is well approximated by hybrid models (H1,H2) which also reproduce the initial larger increase of the \mgii\ blue peak, the absorbed red peak emission in \mgii\ in the later phases of the event (e.g., H1), as well as the small increase of the \mgii\ triplet emission and the small red tail in the \cii\ emission.
        \item {\bf Event 6} (Fig.~\ref{fig_obs}, bottom, and Fig.~\ref{fig_lc_spec2}) is one of the largest events in our sample, with intense, relatively long-lasting ($\sim 1000$~s) and morphologically complex coronal emission. In this event, all the studied footpoints have strong \mgii\ triplet emission, indicating significant presence of NTE, and one of the footpoints (px~1) has \siiv\ emission characterized by strong blueshift and line broadening. The \siiv\ and \mgii\ spectra of px~1 are similar to model E3, though our simulation does not reproduce the unusually broad profiles. The other two footpoints (px~2 and px~3) show no shift (px~2) or modest redshift (px~3) in \siiv\ (see Fig.~\ref{fig_lc_spec2}), and their IRIS spectra are overall similar to model H2.
        \item {\bf Event 8}, although characterized by relatively weak coronal emission (see Table~\ref{table_obs}) shows quite strong TR and chromospheric brightenings (including some \mgii\ triplet emission for px~1), with some complex line profiles (Fig.~\ref{fig_obs8}), somewhat similar to model H2 (Fig.\ref{fig_radyn_1}). Also the spectra of the other two analyzed footpoints (px~2 and px~3) present similarities with the hybrid models (H1 and H2), although with a stronger red component in the \siiv\ emission possibly indicating a stronger contribution of TC compared with NTE for these two footpoints. We also note that this event is the only event in our sample presenting a significant increase in continuum emission in the wavelength range around the \mgii\ triplet (see bottom plots of Fig.~\ref{fig_obs8}). 
        \item {\bf Event 10} presents several peculiarities in its spectral properties (Fig.~\ref{fig_obs10}). The \siiv\ and \cii\ emission before the brightenings is extremely weak (though we note also the shorter exposure time in this dataset). The \siiv\ brightenings are blueshifted, and the chromospheric profiles are highly unusual and show a strong emission increase at shorter wavelengths. These features strongly suggest the presence of highly energetic NTE (inversions with the new IRIS$^2$ code of \cite{SainzDalda2019} also support this conclusion, as we will discuss in detail in a follow-up paper), and it is most similar to model E5. This is also compatible with the very weak overlying coronal emission, because models with highly energetic NTE predict very small coronal emission (because most of the energy is dissipated in the chromosphere; see \citealt{Polito2018} for a more detailed discussion). In fact, although the observed IRIS TR/chromospheric brightenings are associated with heating of overlying small transient loops observed in the AIA 94\AA\ passband (see top right panel of Fig.~\ref{fig_obs10}), the subset of brightenings under the IRIS slit however are at the footpoints of very weakly emitting coronal loops.
    \end{itemize}
\item {\bf Likely presence of NTEs (small \siiv\ blueshifts):}
    \begin{itemize}
        \item {\bf Event 2} (Fig.~\ref{fig_obs2}) involves very short loops, and presents lightcurves suggesting repeated heating, and an overall longer event likely with lower energy flux than assumed in our simulations. The \siiv\ spectrum in Fig.~\ref{fig_obs2} shows a small blueshift suggesting the presence of some NTE. The \mgii\ and \cii\ line profiles show a deep central reversal (not common in our simulations while the nanoflare loop is being heated, in particular for \cii), and the evolution of the \mgii\ emission similar to models H1 (or E4, E6), including the slightly stronger red peak of \cii\ and blue peak of \mgii.
        \item In {\bf event 3} (Fig.~\ref{fig_obs3}) the observed footpoints brightenings are short-lived and intense. The \siiv\ initial emission is generally quite low, its spectral profiles are close to Gaussian and, at least in some cases, show small blueshifts during the brightenings. The chromospheric \cii\ and \mgii\ profiles are often single peaked, which is not easily reproduced by our simulations. For px~1 the observed emission is closest to models E2-E4 (possibly with lower energy flux, considering the generally weaker observed emission), though, as remarked earlier, the close to single peak \cii\ and \mgii\ profiles are not well reproduced. For px~2 the chromospheric emission, including the \cii\ line profiles and modest \mgii\ triplet emission, is overall similar to models E2 and E4, including the stronger blue peak of \cii\ and the small red tail in the \siiv\ profiles.
        \item The footpoints brightenings in {\bf event 5} (Fig.~\ref{fig_obs5}), which has weak and short-lived coronal emission, are barely caught under the slit. The IRIS spectra are characterized by blueshifted \siiv\ emission, barely detected \cii\ emission, unusual \mgii\ profiles, and no significant \mgii\ triplet emission. The \siiv\ blueshifts suggest the presence of some NTE, while the \mgii\ profiles are reminiscent of models C1 and E1, which together with the small intensity increases suggest compatibility with lower energy nanoflares (e.g., see model with total energy of $10^{24}$~erg in \citealt{Polito2018}, which produces blueshifted \siiv\ emission).
    \end{itemize}
\item {\bf Lack of evidence of NTEs:}
    \begin{itemize}
        \item {\bf Event 4} (Fig.~\ref{fig_obs4}), observed in the same AR a few hours later than event 3, is characterized by relatively weak but long-lasting coronal emission. The brightenings observed in the TR and chromosphere are small (especially in \mgii), short-lived, and generally without large \siiv\ Doppler shifts. The \siiv\ spectral profiles of px~1 are similar to model H1 and the later stages (i.e., second heating episode) to model E6, including the red tail of \siiv\ emission, while the single peaked chromospheric profiles are not well matched by our models. For px~2 and px~3 the profiles, including the small blueshift in \siiv\ and the \cii\ (with its brighter red peak) and \mgii\ emission, are similar to models E4-E6.
        \item {\bf Event 7} is a very small event, both in terms of coronal (see Table~\ref{table_obs}) and TR/chromospheric emission (Fig.~\ref{fig_obs7}), with no significant increase in \mgii\ emission. The IRIS spectra are compatible with a very small energy release where the heating transport is dominated by thermal conduction.
        \item {\bf Event 9} is one of the larger events in our sample in terms of its associated coronal emission, and despite that, it does not show significant \mgii\ triplet emission (Fig.~\ref{fig_obs9}). The \siiv\ footpoint brightenings are characterized by either no significant Doppler shift (px~2) or redshift (px~1 and px~3), and the lightcurves generally suggest repeated heating (especially for px~2 and px~3). For px~1 the IRIS spectra are similar to model C1 and E1. For px~2, the brightenings have spectra similar to prediction of models H1-H2 (which however do not reproduce well the deep central reversal observed in the \cii\ spectra of the first brightening, second column of Fig.~\ref{fig_obs9}). The IRIS spectra of the brightening observed in px~3 are more similar to model E1-C1, as in px~1. We note though that the lack of clear evidence of NTEs does not necessarily imply that there are no NTEs. For example, as shown by \cite{Polito2018} (their Fig.~15), for larger energies ($\sim 10^{25}$~erg) also heating by higher energy NTE ($\sim 10$~keV) can produce \siiv\ redshifts.
    \end{itemize}
\end{itemize}

\begin{deluxetable*}{ccccc}
\tablecaption{Observational properties of IRIS spectra and comparison with models, for the events of Table~\ref{table_obs}.}
\tablehead{
\colhead{No.} & \colhead{\siiv\ \tablenotemark{a}} & \colhead{\mgii\ \tablenotemark{b}} & \colhead{comments} & \colhead{closest model(s)} \vspace{-0.3cm} \\ 
\colhead{} & \colhead{Doppler shift} & \colhead{triplet} & \colhead{} & \colhead{} 
}
\startdata
1  & red; blue then red & Y & small \cii\ and \mgii\ central reversal & H1, E2, E3, H2 \\
2  & blue & & deep central reversal & H1 \\
3  & blue; negligible; red &  & small central reversal/single peaked \cii\ and \mgii\ & E2, E6 \\
4  & negligible; blue; red & & small central reversal/single peaked \cii\ and \mgii\ & H1, E4, E6 \\
5  & blue & & unusual \mgii\ profiles, weak \cii\ emission & E1, C1 \\
6  & blue; negligible; red & Y & broad \siiv\ lines & E3, H2 \\
7  & negligible & & no \mgii\ increase & C1 \\
8  & red; blue & Y & continuum enhancement;  & H2, H1 \\
   &  &  & single peaked \cii\ and \mgii\ ; multi-component &  \\
9  & red; neglibile & & unusual (broad, multi-component) profiles & C1, E1; H1, H2 \\
10 & blue & Y & blue components strongly enhanced; weak coronal emission & E5 \\
\enddata
\label{table_obs_mod}
\tablenotetext{a}{The \siiv\ Doppler shifts are relative to the line shift before the heating episode. More than one Doppler shift might be mentioned for each event given the different properties of different footpoints in the same event.}
\tablenotetext{b}{"Y" indicates that \mgii\ triplet emission was observed in at least one footpoint in a given event.}
\end{deluxetable*}

The evidence of accelerated particles in several of these events is a significant indication of magnetic reconnection \citep[e.g.,][]{Cargill2015} as driver of many of these impulsive events. In particular, event 4, 5 and 8, which are analyzed in more detail in \cite{Reale2019a} all present evidence of non-thermal particles. As we discussed in more details in \cite{Reale2019a}, the appearance of the coronal emission and its evolution for several of the events in our sample suggests large-angle reconnection, different from the small angle reconnection driven by loop footpoint shuffling as expected in the classic nanoflare scenario of \cite{Parker1983,Parker1988}, and possibly related to a different driver such as flux emergence \citep[e.g.,][]{Asgari2019}.

%Deconvolution with \iris\ PSF can be important for this type of events, where the emission can be very intense and very localized
% Si IV NEI effects \citep{Kerr2019}

\section{Conclusions}
\label{conclusions}

In this paper we have analyzed a sample of \iris\ observations of moss rapid brightenings associated with coronal loop heating, focusing on the spectral properties of the chromospheric and TR response to the coronal heating.
The events studied here were manually selected to have an evident associated heating of the overlying coronal loops, as observed in the \aia\ 94\AA\ imaging data, and so that at least some of the footpoint brightenings are observed under the \iris\ spectral slit. Typically we find that the coronal temperatures reached by these nanoflare heated loops are up to $\sim 10$MK (see also \citealt{Reale2019a,Reale2019btemp}). The sample we obtained otherwise shows a very broad variety of coronal properties, with loop lengths, and coronal emission and duration spanning more than an order of magnitude (see Table~\ref{table_obs}). Despite this broad range of coronal properties the footpoint brightenings observed with \iris\ have several similarities, including: 
\begin{itemize}
\item duration of the brightenings of less than a minute (with median and average of $\sim 25$~s; ; see Fig.~\ref{fig_hist}), and similar in the TR and chromospheric \iris\ lines, with the exception of the \mgii\ triplet, which is often absent or shorter lived;
\item a large variety of spectral properties (e.g., intensity, Doppler shifts, broadening, line ratios) even for different footpoints in the same coronal event (see e.g., Fig.~\ref{fig_lc_spec1} and~\ref{fig_lc_spec2}); 
\item typically small increases of \siiv\ non-thermal line broadening during the events (see Fig.~\ref{fig_hist}; though there are exceptions, such as e.g., the brightening in the left column of Fig.~\ref{fig_lc_spec2}); 
\item \siiv\ spectral profiles are typically Gaussian, though a few profiles are briefly characterized by multiple components (e.g., see Fig.~\ref{fig_obs8} for event 8 of Table~\ref{table_obs}).
\end{itemize}
We also note that the moss brightenings are often observed at both footpoints (as observed in the \iris\ 1400\AA\ SJI or AIA 1600\AA; see also \citealt{Reale2019a} and \citealt{Testa2014}), though the conjugated footpoint is not observed under the IRIS slit. The simultaneous brightening of both loop footpoints further supports the coronal nanoflare scenario, where magnetic reconnection occurs in the corona and heating is subsequently propagated from the corona to both footpoints simultaneously (at least within the IRIS SJI/AIA cadence), and it seems to be at odds with a scenario in which chromospheric reconnection is the main mechanism responsible for the footpoint brightenings \citep[e.g.,][]{Judge2017}.

The RADYN simulations of nanoflare heated loops we have run with different heating properties (e.g., heating transport -- by TC, NTE, or combination--, duration and total energy of the heating, energy distribution of NTE) reproduce quite well the observed range of intensity, duration, and Doppler shifts of \siiv\ brightenings, as well as chromospheric to TR intensity ratios. These simulations provide a useful guide for the interpretation of the observations (see details in \S\ref{discussion}). We note that even though we mostly find good qualitative agreement between observations and simulations, it is not always possible with the current limited set of simulation parameters to find detailed quantitative agreement for all observed properties. Therefore one of the main conclusions from our analysis is that the combination of all the available spectral information (TR/chromospheric emission), together with the coronal observations, really tightly constrains the heating properties in these impulsive heating events. In Sec.~\ref{discussion} we discussed how the model parameters can be tweaked to improve the agreement between the models and the observations.

Another interesting finding of this work is the powerful diagnostic provided by the \mgii\ triplet emission: our simulations suggests that this emission is caused by heating low in the atmosphere, which, in the scenarios we explored, requires the presence of non-thermal particles (see \S\ref{sims}; however we note that, as we discuss below, we are currently investigating whether the \mgii\ triplet emission might also be caused by \alfven\ wave heating).  Our IRIS observations, for which the \mgii\ triplet is observed in emission in at least 4 out of 10 events, suggest that acceleration of particles takes place, and it is not uncommon, also in smaller -- nanoflare-size -- events, and it is not a prerogative of large flares. Furthermore, we find that evidence of NTE is found also for some of the smaller events (e.g., event 1, Fig.~\ref{fig_lc_spec1}), and, at the same time, some of the brightenings in the larger events of our sample appear compatible with models without NTE.
However we find that only overall larger events (in terms of both coronal and TR/chromospheric emission; e.g., events 6, 8 and 9) are characterized by larger ($\gtrsim 1$) \cii/\siiv\ ratios, likely pointing to higher overall energies, larger duration of the events, more energetic NTE (see Fig.~\ref{fig_hist}, and related text; note also that event 6 and 8 both have \mgii\ triplet emission, indicating the presence of high energy particles). 

Although not all observed spectral properties in a single footpoint brightening are well reproduced by the models we have run, better fits can be found through a more thorough exploration of the parameter space, which will be the focus of follow-up work. The predicted spectral properties in fact critically depend on initial conditions, as well as energy flux and duration of the nanoflare, both of which determine where the energy is deposited. 

The events studied here are heating events in the active region core involving several loops, each heated by a nanoflare size energy release, as evidenced by the good qualitative agreement between the simulations and observations (as well as rough estimates based on imaging data, as in e.g., \citealt{Testa2013}). These type of events have been the focus of earlier studies focused on imaging AIA 94\AA\ observations \citep[e.g.,][]{Testa2012b,UgarteUrra2014,UgarteUrra2017,UgarteUrra2019}. However, the additional spectral diagnostics of our studies (\citealt{Testa2014} and this paper) clearly improves the diagnostic potential to unveil the properties of coronal heating for the hot AR cores. These studies require high spatial and temporal resolution, as provided by IRIS observations, to reveal the highly transient and localized variability at the loop footpoints and therefore, in turn, the impulsive nature of the heating.
We note that, although larger than single nanoflares, the event we analyzed here are nevertheless typically smaller than other small events such as microflares (e.g., \citealt{Warren2016}), which are also visible with hard X-ray spectrometers such as RHESSI \citep{Hannah2008}. Although new hard X-ray instruments have increased sensitivity (e.g., NuSTAR, \citealt{Harrison2013}), their ability to constrain non-thermal emission in very small events such as the ones we focus on here is often limited by the presence of a dominant thermal emission \citep[e.g.,][]{Glesener2017}, as well as by observational limitations due to the use of an astrophysics observatory for solar observations \citep[e.g.,][]{Hannah2019}. We note however that these hard X-ray observations are generally compatible with the NTE distributions suggested by our analysis. For instance, \cite{Wright2017} find for a very small event (characterized by temperature of only $\sim 5$~MK, significantly smaller than in our events) that the NuSTAR spectrum is compatible with  $E_C \lesssim 7$~keV and $\delta \gtrsim 7$ (here we assume $\delta=7$ for all our simulations).

The sample analyzed here is of limited size due to the manual search (which has been restricted to sit-and-stare or few step rasters, and short exposure times) and the general difficulty of finding these footpoint brightenings under the slit. However, we are now working on follow-up work which will allow us to overcome this shortcoming, by using automated detection. \cite{Graham2019} recently presented an algorithm, improved from the one we used on Hi-C data \citep{Testa2013}, which allows us to automatically detect rapid moss variability in \aia\ timeseries. We are currently working on applying this algorithm to \aia\ datacubes co-aligned with \iris\ data (already available on the IRIS data search webpage \footnote{http://iris.lmsal.com/search/ , \iris\ technical note (ITN) 32 https://www.lmsal.com/iris\_science/doc?cmd=dcur\&proj\_num=IS0452\&file\_type=pdf }), to automatically find these brightenings in \iris\ spectral data. This approach will allow us to greatly expand the sample analyzed here and derive more statistically significant properties for this type of events.  The automated approach will also allow to (at least partially) overcome the bias in the selection toward larger events likely affecting the sample in this paper (we manually selected events with relatively high hot coronal emission).

Another issue that we are addressing in follow-up work is the qualitative nature of the comparison we carried out here between observations and RADYN simulations to deduce the heating properties compatible with the observables. We plan to devise a more robust method for this comparison, and the interpretation of the \iris\ observations. In particular, we plan to run RADYN simulations on much finer grid of parameters (as partly discussed in the previous section~\ref{discussion}) exploring a larger parameter space (e.g., initial conditions, duration of the heating, combination of TC and NTE, total nanoflare energy, parameters of NTE), and to find the closest matches to the observed spectra by using an automatic algorithm such as e.g., k-means clustering \citep[e.g.,][]{Panos2018} or using neural networks \citep[e.g.,][]{Osborne2019}.  This will also be complemented by using the IRIS$^2$ inversion code of \cite{SainzDalda2019} for all of our events to obtain further constraints on the initial atmosphere and on the response of the lower atmosphere to small coronal heating events.

Finally, another aspect we are exploring is whether some of these events could be dominated by heating from dissipation of \alfven\ waves rather than non-thermal electrons. Such heating has been implemented in RADYN and used for flare simulations \citep{Kerr2016} and we are now studying \alfven\ wave heating in nanoflare heated loops (Kerr et al., in preparation).  \alfven\ waves are likely to be present, although more difficult than non-thermal particles to constrain observationally, and we want to investigate whether they have non-negligible effect on coronal heating in active region cores. Preliminary results of our RADYN simulations seem to indicate that impulsive \alfven\ wave heating cannot easily reproduce TR redshifts (e.g., in IRIS \siiv\ lines), as opposed to NTE heating which can reproduce both blue- and redshifts, and this therefore suggests they are likely not the sole heating transport process at work in these impulsive heating events.  

To conclude, in this work we have demonstrated the diagnostic capabilities of IRIS that by observing lines formed over different atmospheric layers provides tight constraints on the energy deposition in nanoflares.

\acknowledgements
We thank the referee for their careful reading of this paper and the very useful comments which greatly helped improving the paper.
PT and VP were funded for this work by the NASA Heliophysics Guest Investigator grant NNX15AF50G. P.T. was also supported by contracts 8100002705 and SP02H1701R from Lockheed-Martin to SAO, and NASA contract NNM07AB07C to the Smithsonian Astrophysical Observatory.  B.D.P. was supported by NASA contract NNG09FA40C (IRIS).
This work has benefited from discussions at the International Space Science Institute (ISSI) meetings on "New Diagnostics of Particle Acceleration in Solar Coronal Nanoflares from Chromospheric Observations and Modeling", and "Heating in the magnetic chromosphere" where topics relevant to this work were discussed with other colleagues. 
IRIS is a NASA small explorer mission developed and operated by LMSAL with mission operations executed at NASA Ames Research center and major
contributions to downlink communications funded by ESA and the Norwegian Space Centre.
This research has made use of NASA's Astrophysics Data System and of the SolarSoft package for IDL.
 
 \appendix
 
 \section{Additional datasets}
 \label{appendix:data}
\begin{figure}
	\includegraphics[height=5.5cm]{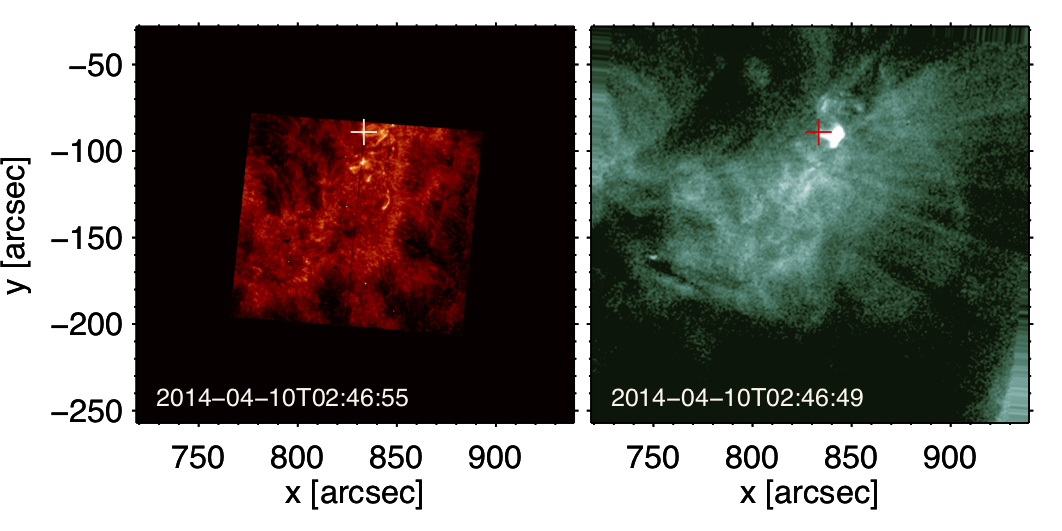}	
  	\vspace{-0.2cm}
         \includegraphics[height=5.8cm]{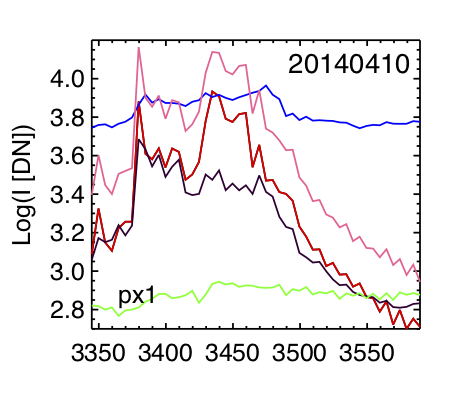}
  	\vspace{-0.2cm}
         \includegraphics[width=18cm]{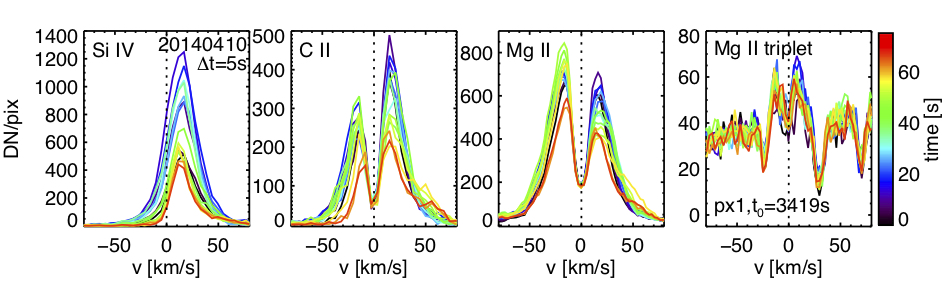}
	\caption{\iris\ 1400\AA\ slit-jaw images ({\em top left}) and \aia\ 94\AA\ images ({\em top right}) showing a moss brightening event observed on 2014-04-10 (event 2 in Table~\ref{table_obs}), and associated hot loop emission. For the pixel marked in the images, we show the temporal evolution of \iris\ spectral observables (lightcurves and spectra of different spectral lines). Figures are in the same format as Fig.\ref{fig_obs} and ~\ref{fig_lc_spec1}.}
	\label{fig_obs2}
\end{figure}

\begin{figure*}
	\centering
	\hspace{-1.3cm} \vspace{-0.2cm}
	\includegraphics[width=13cm]{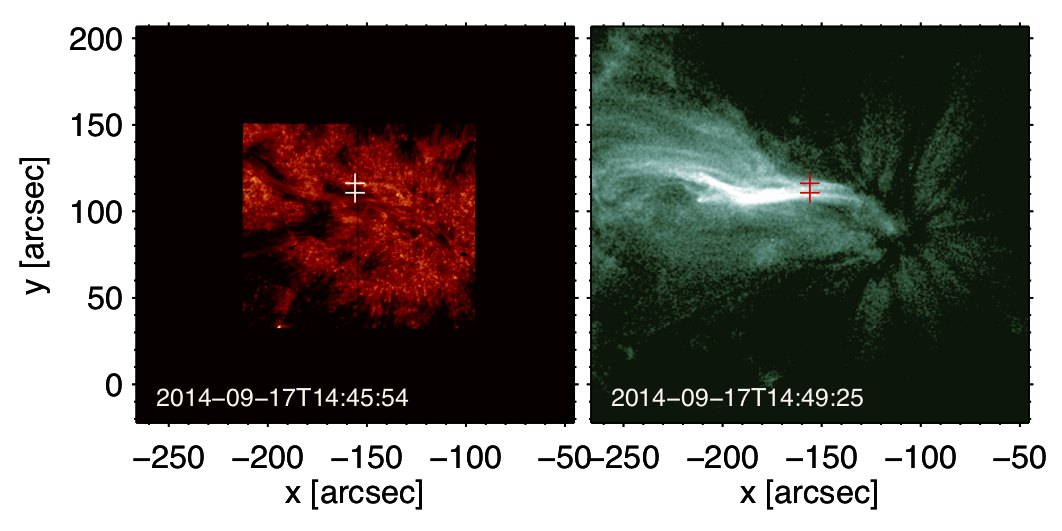}	
  	\vspace{-0.2cm}
         \includegraphics[width=13cm]{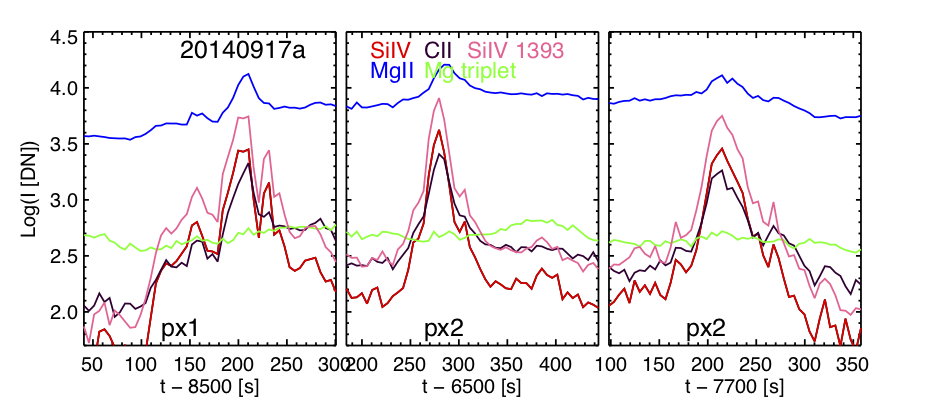}
  	\vspace{-0.2cm}
         \includegraphics[width=13cm]{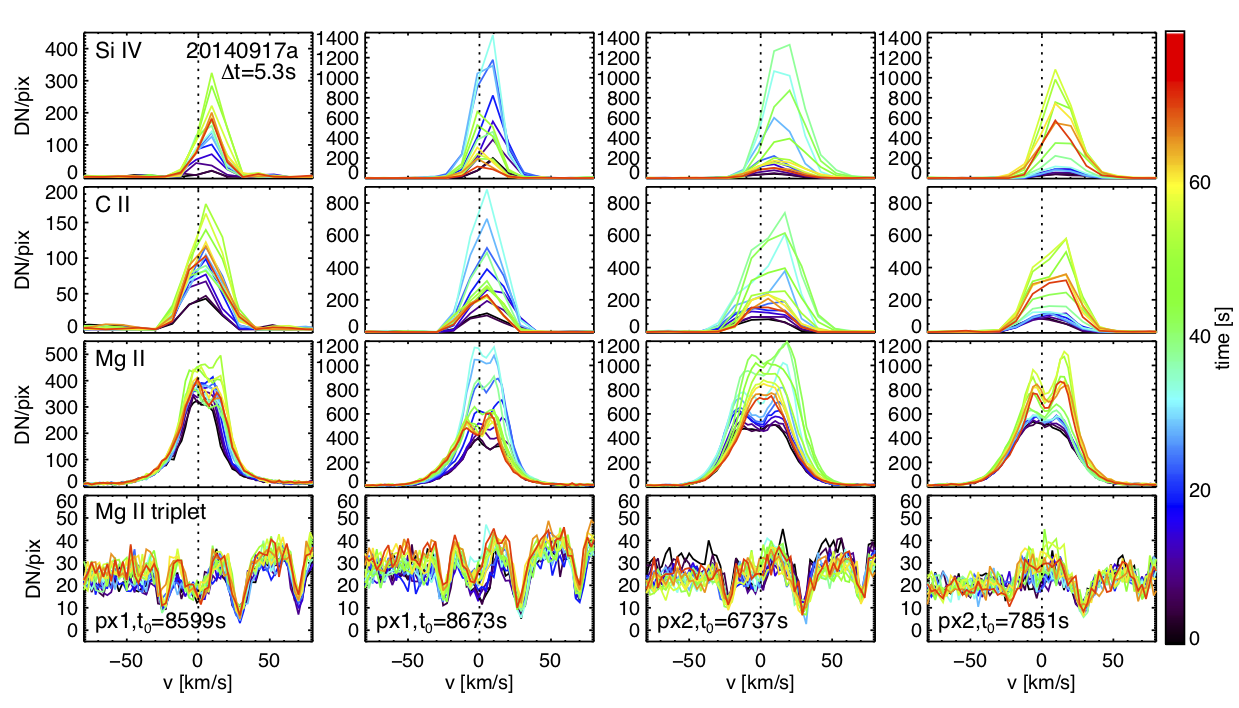}
	\caption{\iris\ 1400\AA\ slit-jaw images ({\em top left}) and \aia\ 94\AA\ images ({\em top right}) showing moss brightenings observed on 2014-09-17 (event 3 in Table~\ref{table_obs}), and associated hot loop emission. For the 3 pixels marked in the images, we show (ordered by increasing value of solar y) the temporal evolution of \iris\ spectral observables (lightcurves -- {\em second row} -- and spectra -- {\em bottom four rows} --  of different spectral lines). Figures are in the same format as Fig.\ref{fig_obs} and ~\ref{fig_lc_spec1}.}
	\label{fig_obs3}
\end{figure*}

\begin{figure*}
	\centering
	\hspace{-1.3cm}\vspace{-0.3cm}
	\includegraphics[width=13cm]{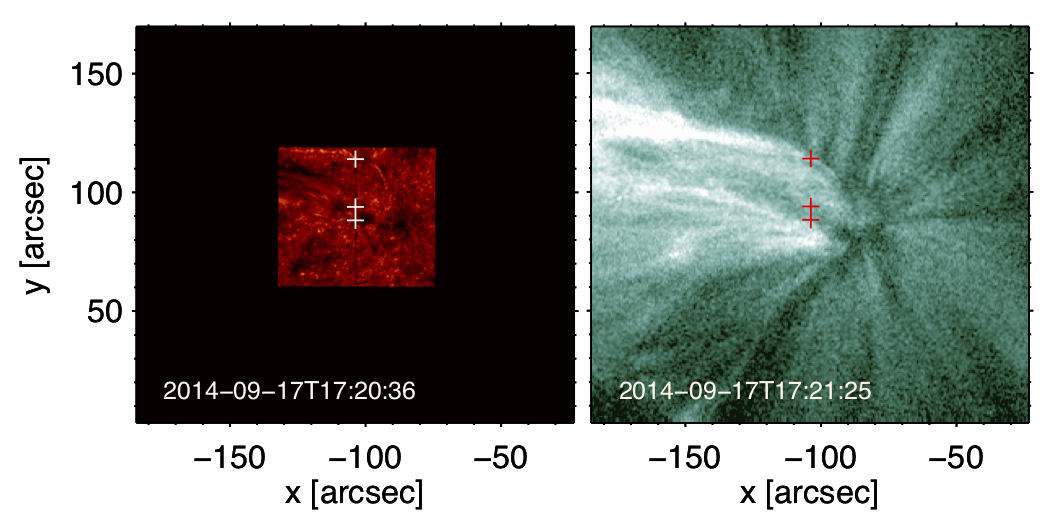}	
  	\vspace{-0.2cm}
        \includegraphics[width=13cm]{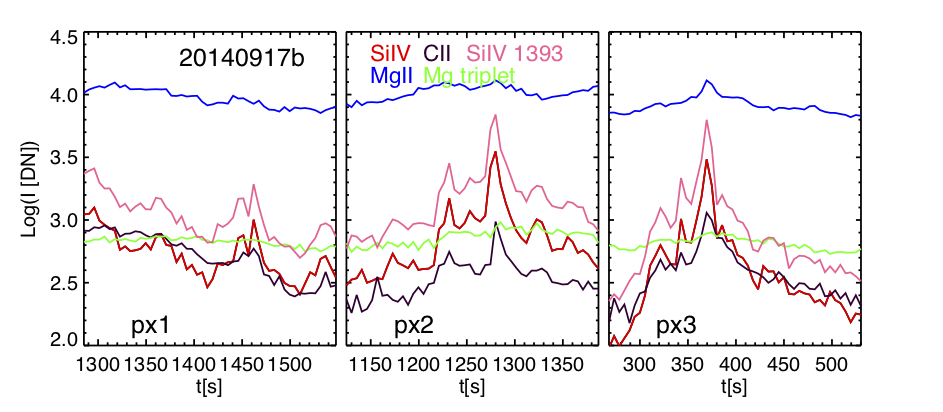}
  	\vspace{-0.3cm}
         \includegraphics[width=13cm]{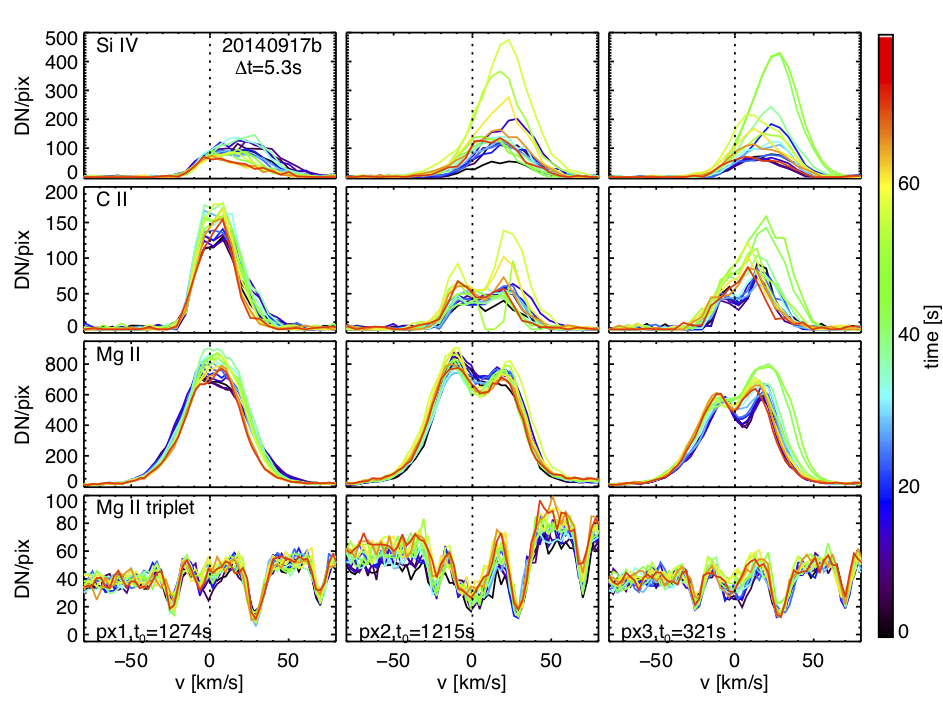}
	\caption{\iris\ 1400\AA\ slit-jaw images ({\em top left}) and \aia\ 94\AA\ images ({\em top right}) showing moss brightenings observed on 2014-09-17 (event 4 in Table~\ref{table_obs}; in the same AR, and a few hours later, as event 3 shown in Fig.~\ref{fig_obs3}), and associated hot loop emission. For the 3 pixels marked in the images, we show (ordered by increasing value of solar y) the temporal evolution of \iris\ spectral observables (lightcurves -- {\em second row} -- and spectra -- {\em bottom four rows} --  of different spectral lines). Figures are in the same format as Fig.\ref{fig_obs} and ~\ref{fig_lc_spec1}.}
	\label{fig_obs4}
\end{figure*}

\begin{figure*}
	\centering
	\hspace{-1.6cm} \vspace{-0.35cm}
	\includegraphics[width=10.5cm]{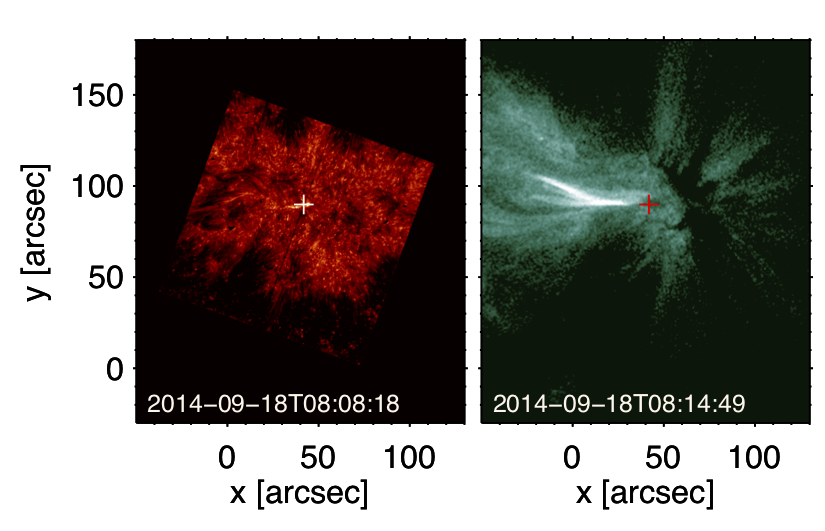}	
 	\vspace{-0.35cm}
        \includegraphics[width=11cm]{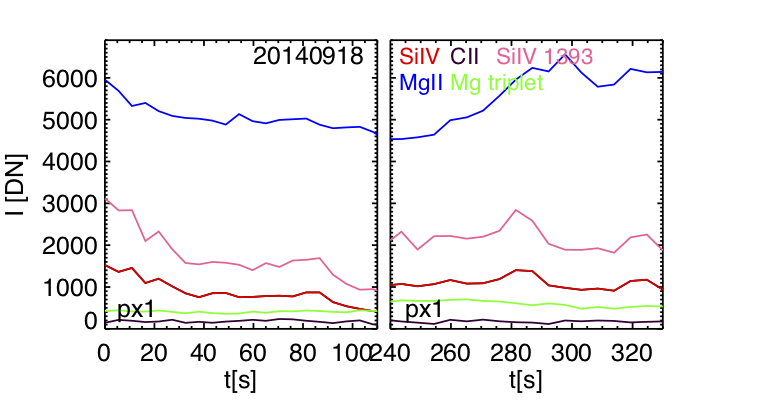}
 	\vspace{-0.3cm}
         \includegraphics[width=12cm]{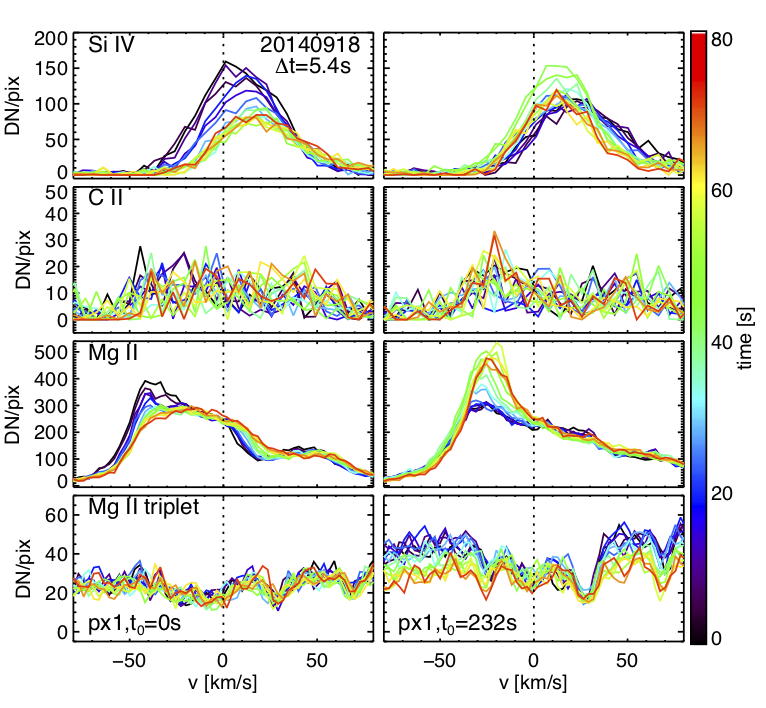}
	\caption{\iris\ 1400\AA\ slit-jaw images ({\em top left}) and \aia\ 94\AA\ images ({\em top right}) showing moss brightenings event observed on 2014-09-18 (event 5 in Table~\ref{table_obs}), and associated hot loop emission. For the pixel marked in the images, we show (for 2 moss brightening events occurring at different times) the temporal evolution of \iris\ spectral observables (lightcurves -- {\em second row} -- and spectra -- {\em bottom four rows} --  of different spectral lines). Figures are in the same format as Fig.\ref{fig_obs} and ~\ref{fig_lc_spec1}.}
	\label{fig_obs5}
\end{figure*}

\begin{figure*}
	\includegraphics[width=10cm]{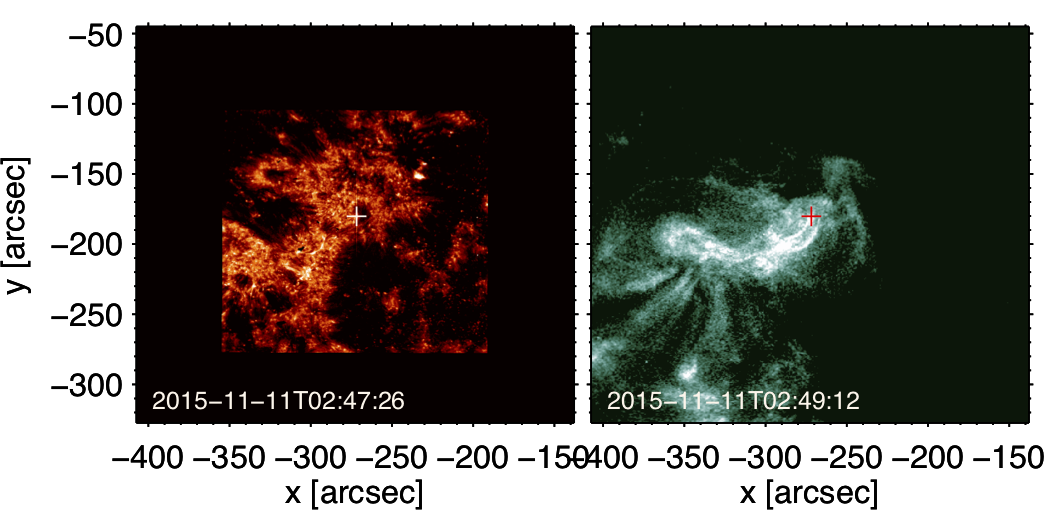} 
 	\vspace{-0.25cm}
         \includegraphics[width=8cm]{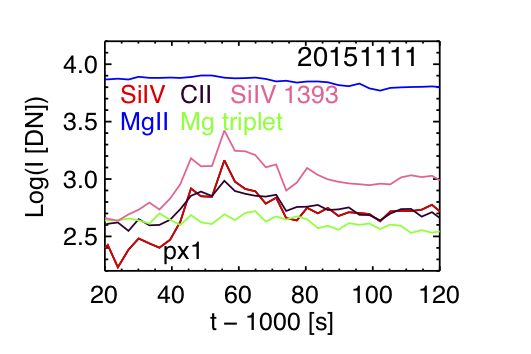} 
         \includegraphics[width=18cm]{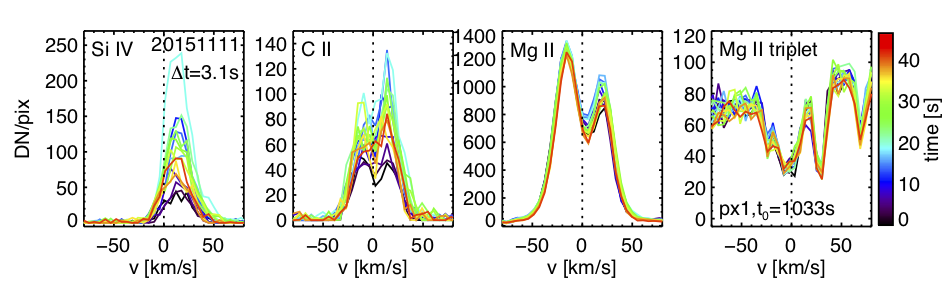}
	\caption{\iris\ 1400\AA\ slit-jaw images ({\em top left}) and \aia\ 94\AA\ images ({\em top right}) showing moss brightenings event observed on 2015-11-11 (event 7 in Table~\ref{table_obs}), and associated hot loop emission. For the pixel marked in the images, we show the temporal evolution of \iris\ spectral observables (lightcurves -- {\em second row} -- and spectra -- {\em bottom four rows} --  of different spectral lines). Figures are in the same format as Fig.\ref{fig_obs} and ~\ref{fig_lc_spec1}.}
	\label{fig_obs7}
\end{figure*}

\begin{figure*}
	\centering
	\hspace{-1.2cm}  	\vspace{-0.3cm}
	\includegraphics[width=13cm]{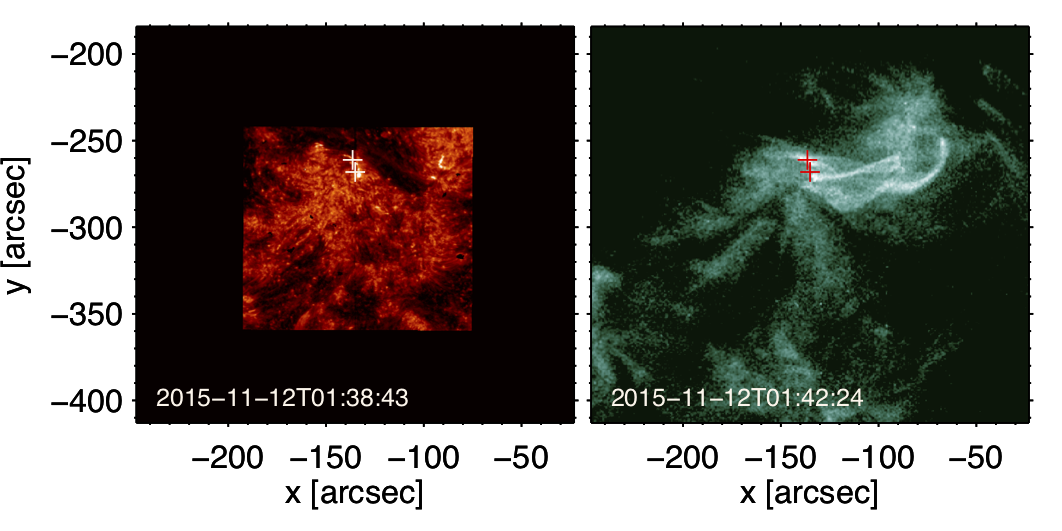}	
 	\vspace{-0.4cm}
         \includegraphics[width=14cm]{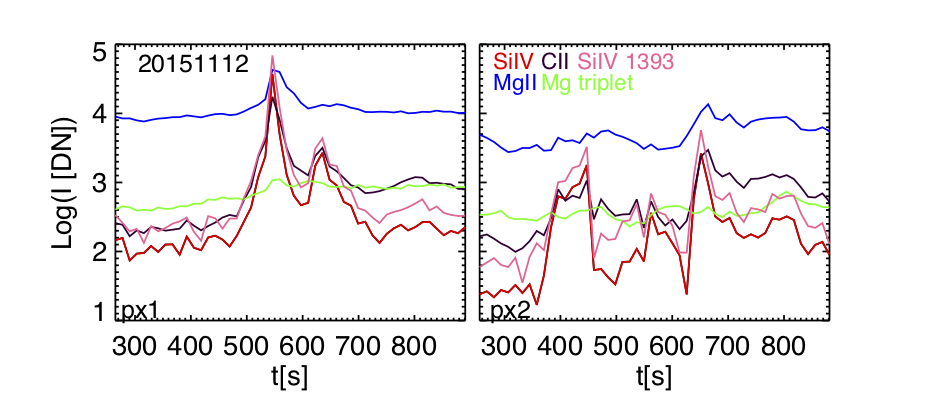}
 	\vspace{-0.2cm}
         \includegraphics[width=13cm]{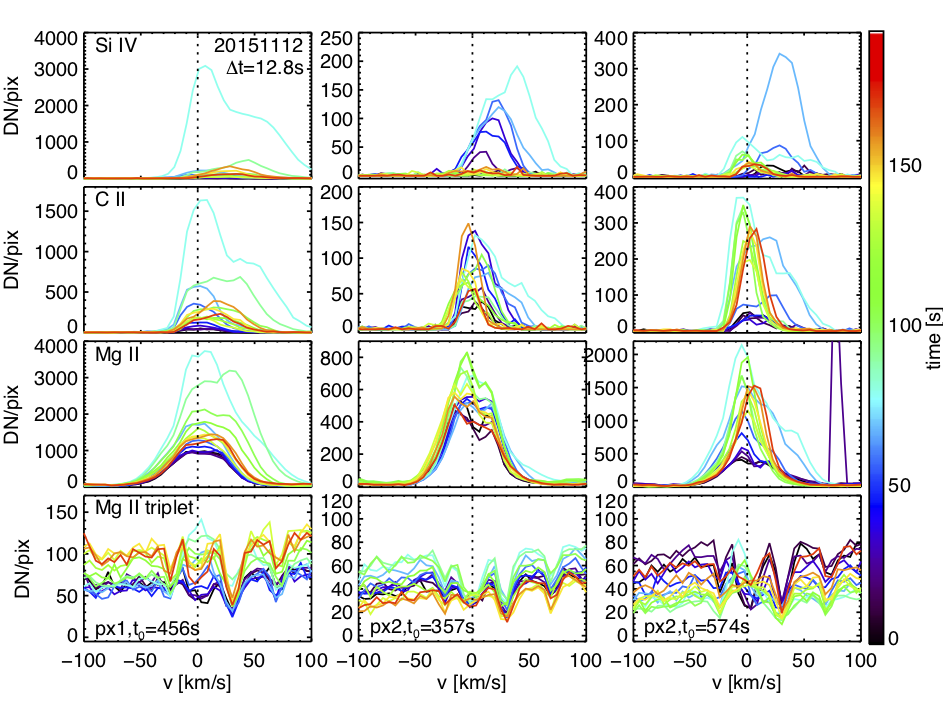}
	\caption{\iris\ 1400\AA\ slit-jaw images ({\em top left}) and \aia\ 94\AA\ images ({\em top right}) showing moss brightenings observed on 2015-11-12 (event 8 in Table~\ref{table_obs}), and associated hot loop emission. For the 2 pixels marked in the images, we show (ordered by increasing value of solar y, and time of occurrence of event) the temporal evolution of \iris\ spectral observables (lightcurves -- {\em second row} -- and spectra -- {\em bottom four rows} --  of different spectral lines). Figures are in the same format as Fig.\ref{fig_obs} and ~\ref{fig_lc_spec1}.}
	\label{fig_obs8}
\end{figure*}

\begin{figure*}
	\centering
	\hspace{-1cm} 	\vspace{-0.3cm}
    \includegraphics[width=13cm]{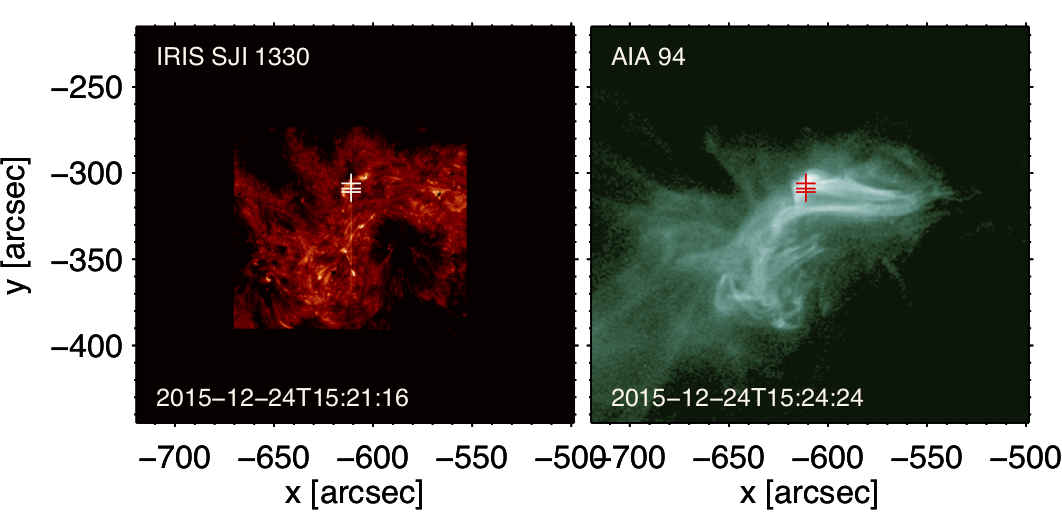}	
 	\vspace{-0.2cm}
         \includegraphics[width=13cm]{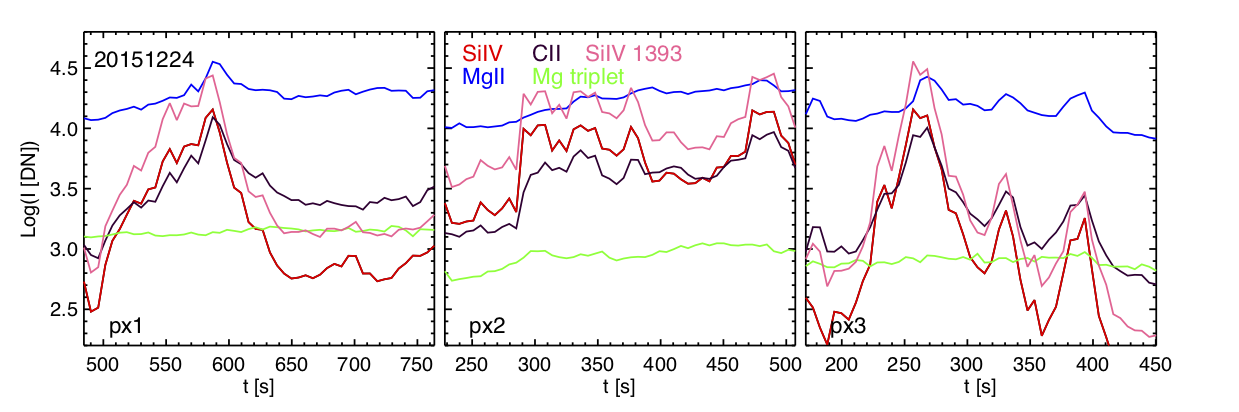}
 	\vspace{-0.2cm}
         \includegraphics[width=13cm]{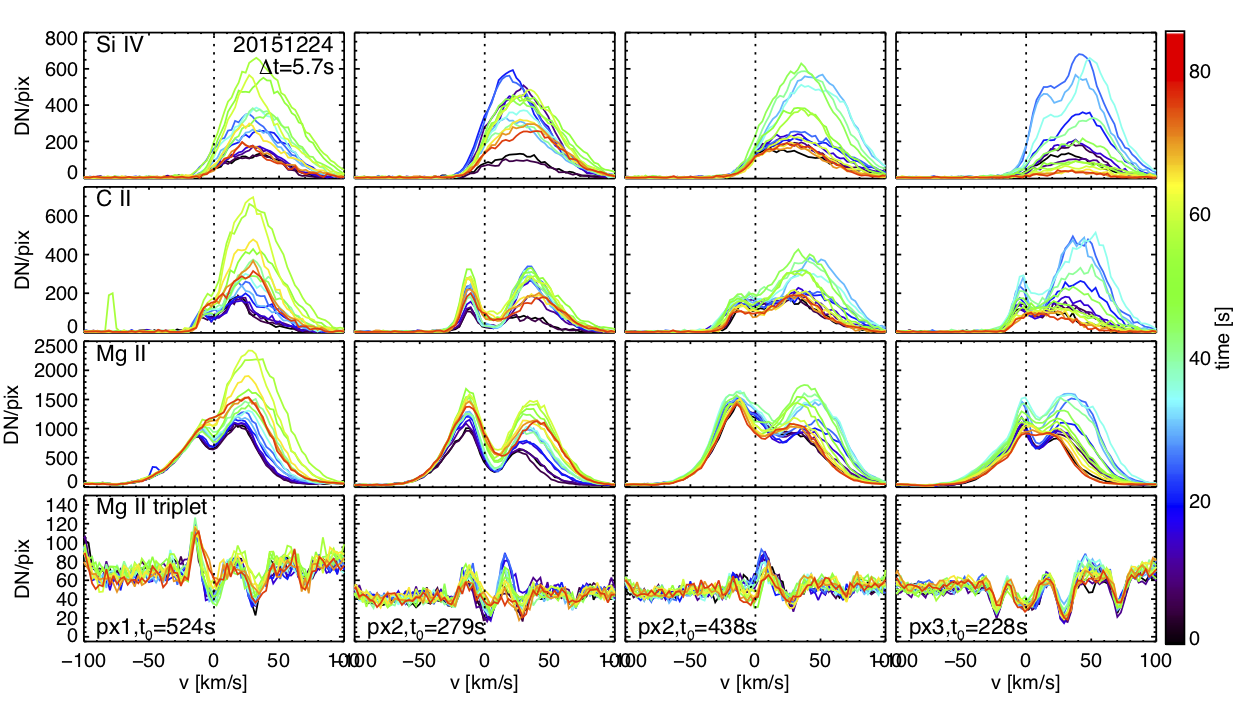}
	\caption{\iris\ 1400\AA\ slit-jaw images ({\em top left}) and \aia\ 94\AA\ images ({\em top right}) showing moss brightenings observed on 2015-12-24 (event 9 in Table~\ref{table_obs}), and associated hot loop emission. For the 3 pixels marked in the images, we show (ordered by increasing value of solar y) the temporal evolution of \iris\ spectral observables (lightcurves -- {\em second row} -- and spectra -- {\em bottom four rows} --  of different spectral lines). Figures are in the same format as Fig.\ref{fig_obs} and ~\ref{fig_lc_spec1}.}
	\label{fig_obs9}
\end{figure*}

\begin{figure*}
	\centering
	\hspace{-1cm}
	\includegraphics[width=12cm]{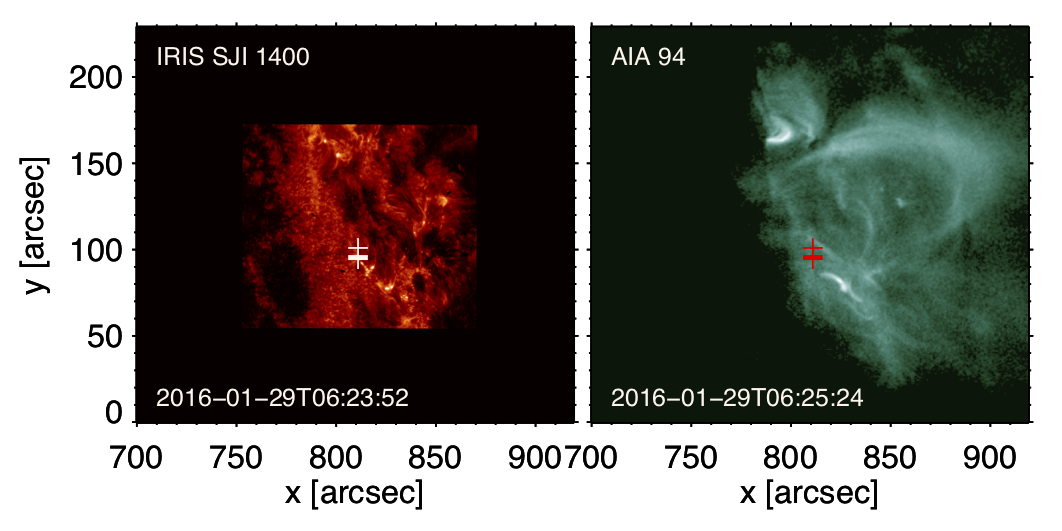}	
         \includegraphics[width=13cm]{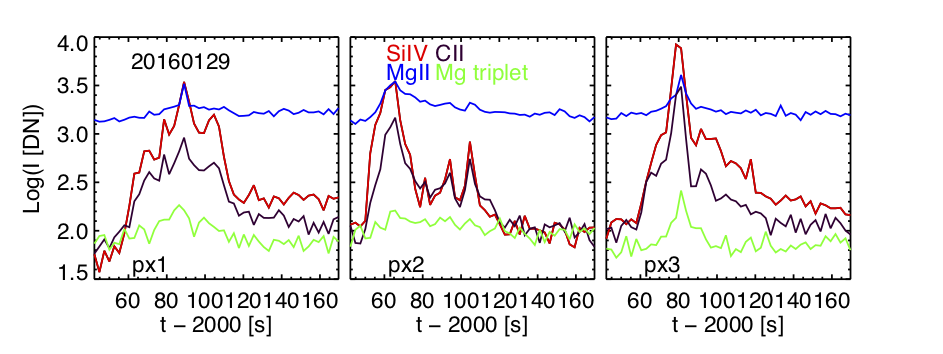}
         \includegraphics[width=13cm]{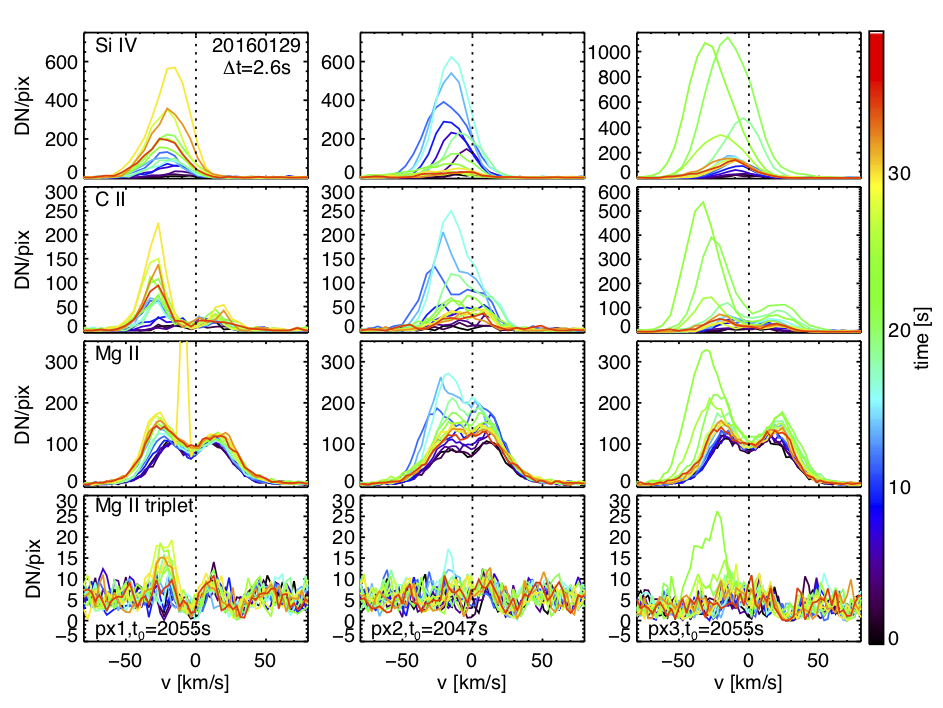}
	\caption{\iris\ 1400\AA\ slit-jaw images ({\em top left}) and \aia\ 94\AA\ images ({\em top right}) showing moss brightenings observed on 2016-01-29 (event 10 in Table~\ref{table_obs}), and associated hot loop emission. For the 3 pixels marked in the images, we show (ordered by increasing value of solar y) the temporal evolution of \iris\ spectral observables (lightcurves -- {\em second row} -- and spectra -- {\em bottom four rows} --  of different spectral lines). Figures are in the same format as Fig.\ref{fig_obs} and ~\ref{fig_lc_spec1}.}
	\label{fig_obs10}
\end{figure*}

 In this Appendix we present imaging and spectral observations for the other events not shown in the main text (in particular event 2-5 and 7-10 of Table~\ref{table_obs}). In Figures~\ref{fig_obs2}-\ref{fig_obs10}, for each event we present \iris\ 1400\AA\ slit-jaw images and \aia\ 94\AA\ images in the same format as Figure~\ref{fig_obs}, to show the location of the loop footpoints where IRIS spectra have been analyzed, and the morphology of the coronal emission. In the same figures we also show lightcurves and evolution of IRIS spectra during the brightenings in the same format as in Figures~\ref{fig_lc_spec1} and~\ref{fig_lc_spec2}.  In \S\ref{discussion}, these datasets are discussed in detail, and compared with IRIS synthetic spectra from 1D RADYN simulations of nanoflare heated loops, to infer the properties of the heating.
 
 % The bibliography
%\bibliography{bib_stars}

\end{document}